\documentclass[12pt,usenames,dvipsnames]{article}

\usepackage{latexsym}
\usepackage{amssymb,amsfonts,amsmath}
\usepackage{graphicx} 
\usepackage{indentfirst}
\usepackage{bbm}
\usepackage{amssymb}
\usepackage{verbatim}
\usepackage{amsmath, amsthm,amssymb}
\usepackage{mathrsfs}
\usepackage{hyperref}
\usepackage{amsfonts}
\usepackage{dsfont}
\usepackage{cite}
\usepackage{xcolor}
\usepackage{enumerate}
\usepackage{cleveref}

\parskip=\medskipamount

\newcommand {\cD}{{\cal D}}
\newcommand {\cE}{{\cal E}}

\newcommand {\cH}{{\cal H}}

\newcommand {\cL}{{\cal L}}

\newcommand {\cN}{{\cal N}}

\def\a{\alpha}
\def \bi{\bibitem}

\def\b{\beta}

\def\d{\delta}
\def\e{\epsilon}
\def\f{\phi}
\def\g{\gamma}
\def\G{\Gamma}

\def\l{\lambda}
\def\m{\mu}

\def\o{\omega}
\def\p{\pi}
\def\q{\theta}
\def\r{\rho}
\def\s{\sigma}

\def\z{\zeta}
\def\D{\Delta}
\def\F{\Phi}
\def\J{\Psi}
\def\L{\Lambda}
\def\O{\Omega}

\def\U{\Upsilon}
\def\X{\Xi}
\def\tr{{\rm tr}}
\def\rd{{\rm d}}
\def\ri{{\rm i}}
\def\re{{\rm e}}

\newcommand{\ad}{{\dot{\alpha}}}                           %
\newcommand{\bd}{{\dot{\beta}}}                            %
\newcommand{\ve}{\varepsilon}                            %
\newcommand{\cDB}{{\bar\cD}}                            %

\renewcommand{\aa}{{\a\ad}}
\newcommand{\bb}{{\b\bd}}
\newcommand{\pa}{\partial}                           %
\newcommand{\hf}{\frac12}

\newcommand{\vf}{\varphi}

\newcommand{\be}{\begin{equation}}
\newcommand{\ee}{\end{equation}}
\newcommand{\bea}{\begin{eqnarray}}
\newcommand{\eea}{\end{eqnarray}}
\newcommand{\non}{\nonumber}

\newcommand{\bm}[1]{\mbox{\boldmath$#1$}}

\def\double #1{#1{\hbox{\kern-2pt $#1$}}}

\newcommand{\gd}{{\dot\g}}
\newcommand{\dd}{{\dot\d}}

\newif\ifdtup

\newcommand{\bsubeq}{\begin{subequations}}
\newcommand{\esubeq}{\end{subequations}}

\numberwithin{equation}{section}

\newcommand{\sSU}{\mathsf{SU}}

\newcommand{\sU}{\mathsf{U}}

\begin{document}

\begin{titlepage}
\begin{flushright}
June, 2024 \\
\end{flushright}
\vspace{5mm}

\begin{center}
{\Large \bf 
Covariant quantisation of tensor multiplet models 
}
\end{center}

\begin{center}

{\bf Sergei M. Kuzenko and Emmanouil S. N. Raptakis} \\
\vspace{5mm}

\footnotesize{
{\it Department of Physics M013, The University of Western Australia\\
35 Stirling Highway, Perth W.A. 6009, Australia}}  
~\\
\vspace{2mm}
~\\
Email: \texttt{ 
sergei.kuzenko@uwa.edu.au, emmanouil.raptakis@uwa.edu.au}\\
\vspace{2mm}

\end{center}

\begin{abstract}
\baselineskip=14pt
The Batalin-Vilkovisky formalism is applied to quantise the ${\cal N}=1$ supersymmetric generalisation of the Freedman-Townsend (FT) model, which was proposed by Lindstr\"om and Ro\v{c}ek in 1983 in Minkowski superspace and is lifted to a supergravity background in this paper. This super FT theory describes a non-Abelian tensor multiplet and is known to be classically equivalent to a supersymmetric nonlinear sigma model. Using path integral considerations, we demonstrate that this equivalence holds at the quantum level in the sense that the quantum supercurrents in the two theories coincide.  A modified Faddeev-Popov procedure is employed to quantise models for the ${\cal N}=2$ tensor multiplet in harmonic superspace. The obtained results agree with those derived by applying the Batalin-Vilkovisky scheme within the harmonic superspace setting.
\end{abstract}
\vspace{5mm}

\vfill

\vfill
\end{titlepage}

\newpage
\renewcommand{\thefootnote}{\arabic{footnote}}
\setcounter{footnote}{0}

\tableofcontents{}
\vspace{1cm}
\bigskip\hrule

\allowdisplaybreaks

\section{Introduction}

With the discovery of simple supergravity in four dimensions \cite{FvNF,DZ} 
and its extended and higher-dimensional generalisations, new types of gauge theories 
(as compared with the standard Yang-Mills theories) have been introduced in quantum field theory. Their specific new features in the Lagrangian formalism are: (i) open gauge algebra; and/or (ii) linearly dependent gauge generators. These imply that covariant quantisation of such theories cannot be carried out using the standard Faddeev-Popov approach \cite{Faddeev:1967fc}. Methods for covariant quantisation of gauge theories with open gauge algebra were developed by Kallosh \cite{Kallosh:1977ik, Kallosh:1978de} and de Wit and van Holten \cite{deWit:1978hyh}.
Quantisation of a general irreducible gauge with open algebra was accomplished by Batalin and Vilkovisky \cite{Batalin:1981jr}. Reducible gauge theories (also known as  gauge theories with linearly dependent generators) naturally arise when one deals with a gauge two-form in four dimensions \cite{OP,KR,CS} and, more generally, gauge $p$-forms in diverse dimensions. These naturally occur in supergravity theories, see, e.g., \cite{GSO,CSF,CJS, Cremmer:1979up} for early publications and \cite{VanN,SalamSezgin,Tanii} for reviews. Several consistent quantisation procedures have been developed to quantise reducible Abelian  gauge theories such as gauge $p$-forms \cite{Schwarz1,Schwarz2,Siegel, Thierry-Mieg:1980ihu, Obukhov,BK88}, including the formulations proposed in \cite{Siegel,BK88} which apply in the supersymmetric case.  

The procedures developed in \cite{Schwarz1,Schwarz2,Siegel, Thierry-Mieg:1980ihu, Obukhov,BK88}
are not directly applicable for the quantisation of the model for a non-Abelian antisymmetric tensor gauge field proposed by Freedman and Townsend \cite{FT}.\footnote{This work built on two earlier unpublished papers
\cite{Freedman:1977pa, Townsend:1979ff}.} 
An important property of the Freedman-Townsend (FT) theory is its classical equivalence to the principal chiral nonlinear sigma model. The same model had also been formulated by Thierry-Mieg in 1980 \cite{Thierry-Mieg:1980ihu}, however his work was published only ten years later.\footnote{One of us (SMK) is grateful to Jean Thierry-Mieg for sharing the story of his work \cite{Thierry-Mieg:1980ihu}.} 
Thierry-Mieg's work also built on the unpublished papers by Freedman
\cite{Freedman:1977pa} and Townsend \cite{Townsend:1979ff}.

An ideal procedure to quantise the FT model is the Batalin-Vilkovisky (BV) formalism 
\cite{BV}, which is universally applicable to general reducible gauge theories with open algebra. 
Following earlier attempts to quantise the FT model \cite{Thierry-Mieg:1982eby, deAlwis:1987fr} using generalised Faddeev-Popov and BRST techniques,
the BV quantisation of the FT model was carried out in 
\cite{BK87, deAlwis:1987sr, Batlle:1988rd} (see \cite{Gomis:1994he} for a review).\footnote{Ref. \cite{BK87}, completed in August 1986, was one of the earliest applications of the BV formalism. It was accepted for publication in Sov. J. Nucl. Phys. in 1987. Shortly before publication it was withdrawn, after the authors had been informed by a colleague that the same problem had already been solved elsewhere. Due to a limited access to the journals, at the time it was not possible  to verify this information, which in fact turned out  to be false.}
Quantum equivalence of the FT model and the principal chiral nonlinear $\sigma$-model was studied in \cite{BK87, deAlwis:1987sr, FradkinT, Nielsen:1990hv}. In the Hamiltonian approach, quantisation of the FT model and its quantum equivalence to the $\s$-model was studied by Frolov and Slavnov \cite{Slavnov:1988sj}.

The $\cN=1$ supersymmetric extension of the FT model was introduced by Lindstr\"om and Ro\v{c}ek in 1983 \cite{LR83}.\footnote{Actually, the authors of \cite{LR83} gave credit to Cremmer and Ferrara, and Gates and Siegel for this model. } 
Further studies and generalisations of this theory have been given in \cite{CLL, Brandt:1998pp, Furuta:2001kx}. This is a non-Abelian extension of the $\cN=1$ tensor multiplet proposed by Siegel 
\cite{Siegel-tensor}. 
In this paper we will couple the $\cN=1$ super FT model to supergravity. 

Covariant quantisation of the free tensor multiplet model in a supergravity background was carried out in \cite{BK88,GNSZ}. The latter theory is classically equivalent to a non-conformal scalar multiplet model \cite{Siegel-tensor}. Their quantum equivalence was established in \cite{GNSZ} in the case of  an on-shell 
supergravity background, and in \cite{BK88} for an arbitrary supergravity background. 

In this paper we make use of the Batalin-Vilkovisky procedure to quantise the $\cN=1$ super 
FT model, which describes a non-Abelian tensor multiplet. 
At the classical level, this theory is known to be equivalent to a supersymmetric nonlinear sigma model \cite{CLL}. 
Using path integral considerations, we demonstrate that this equivalence holds at the quantum level. 

To the best of our knowledge, an $\cN=2$ supersymmetric extension of the FT model 
is not known in the literature. In this paper, a modified Faddeev-Popov procedure is employed to quantise models for the (Abelian) ${\cal N}=2$ tensor multiplet in harmonic superspace.

This paper is organised as follows. Section \ref{section2} is devoted to: (i) coupling the super FT model to old minimal supergravity; (ii) carrying out BV quantisation of the $\cN=1$ super FT model  \eqref{N=1FT}; and (iii) proving its quantum equivalence to the locally supersymmetric nonlinear $\sigma$-model \eqref{NLSM}. Section \ref{section3} is devoted to covariant quantisation of $\cN=2$ tensor multiplet models. For this goal we we make use of the modified Faddeev-Popov procedure described in \cite{BK88}. Concluding comments are given in section \ref{section4}. Rudiments of harmonic superspace are collected in appendix \ref{appendixA}. BV quantisation of the $\cN=2$ Abelian vector and tensor multiplets is carried out in appendix \ref{appendixB}.


\section{Supersymmetric Freedman-Townsend model}
\label{section2}

The $\cN=1$ supersymmetric extension \cite{LR83} of the FT model is 
described by the action\footnote{Our superspace notation and conventions mostly correspond to the textbooks \cite{WB, Buchbinder:1998qv}. We also make use of the notational shorthands $\rd^{4|4}z :=  \rd^4 x \rd^2 \q \rd^2 \bar \q $, $\rd^{4|2}z_+ := \rd^4 x \rd^2 \q$ and $\rd^{4|2}z_- := \rd^4 x \rd^2 \bar{\q}$ to denote the integration measures of the Minkowski superspace and its chiral and antichiral subspaces.}
\begin{align}
	\label{N=1FT}
	S_0[\chi,\bar{\chi},V] = \hf \text{tr} \int \rd^{4|2}z_+ \, \mathcal{E} \, \chi^\a W_\a + \hf \text{tr} \int \rd^{4|2}z_- \, \bar{\mathcal{E}} \, \bar{\chi}_\ad \bar{W}^\ad + \hf  \text{tr} \int \rd^{4|4} z \, E \, V^2~,
\end{align}
where $\chi_\a$, $W_\a$ and $V = {V}^\dagger$ are Lie algebra-valued superfields\footnote{We will say that $\U$ is a Lie algebra-valued superfield if $\U = \U^I T^I$, where $\U^I$ is a superfield and $T^I$ are the generators of a semi-simple Lie group $G$. The latter are normalised as $\text{tr}(T^I T^J) = \d^{IJ}$ and satisfy the commutation relations $[T^I , T^J] = \ri f^{IJK}T^K$, where $f^{IJK}$ are the totally antisymmetric structure constants.
} 
satisfying:
\begin{align}
	\label{2.2}
	\bar{\nabla}_\ad \chi_\a = 0 ~, \qquad \bar{\nabla}_{\ad} W_\a = 0~, \qquad \nabla^\a W_\a = \bar{\nabla}_\ad \bar{W}^\ad ~,
\end{align}
and $E^{-1}= {\rm Ber} (E_A{}^M)$ is the full superspace measure, while $\mathcal{E}$ denotes the chiral density.
Operators $\nabla_A = (\nabla_a,\nabla_\a,\bar{\nabla}^\ad)$ above are gauge-covariant derivatives, 
$\nabla_A = E_A{}^M \pa_M + \hf \O_A{}^{bc} M_{bc} + \ri \,\G_A{}^I T^I$, with $M_{bc} $ the Lorentz generators. The covariant derivatives obey the algebra:
\begin{subequations}
	\label{SYMAlgebra}
	\begin{align}
		\{ \nabla_\a , \nabla_\b \} &= - 4 \bar{R} M_{\a \b} ~, \quad \{ \nabla_\a , \bar{\nabla}_\ad \} = - 2 \ri \nabla_\aa ~, \quad \{ \bar{\nabla}_\ad , \bar{\nabla}_\bd \} = 4 R \bar{M}_{\ad \bd} ~, \\
		[\bar{\nabla}_\ad , \nabla_\bb ] &=  -\ri \ve_{\ad \bd} \Big( R \nabla_\b + G_\b{}^{\gd} \bar{\nabla}_\gd - \bar{\nabla}^\gd G_\b{}^{\dd} \bar{M}_{\gd \dd} + 2 W_\b{}^{\g \d} M_{\g \d} - {W}_\b \Big) - \ri \nabla_{\b} R \bar{M}_{\ad \bd} ~, \\
		[{\nabla}_\a , \nabla_\bb ] &= \ri \ve_{\a \b} \Big( \bar{R} \bar{\nabla}_\bd + G^\g{}_{\bd} {\nabla}_\g - {\nabla}^\g G^{\d}{}_\bd \bar{M}_{\gd \dd} \bar{M}_{\gd \dd} + 2 \bar{W}_\bd{}^{\gd \dd} \bar{M}_{\gd \dd} + \bar{W}_\bd \Big) + \ri \bar{\nabla}_{\bd} \bar{R} \bar{M}_{\ad \bd} ~.
	\end{align}
\end{subequations}
Here the torsion tensors $R$, $G_a = {\bar G}_a$ and
$W_{\a \b \g} = W_{(\a \b\g)}$ satisfy the  Bianchi identities:
\begin{subequations}
	\bea
	&\bar{\nabla}_\ad R= 0~,~~~~~~\bar{\nabla}_\ad W_{\a \b \g} = 0~,
	\\
	&
	\bar{\nabla}^\gd G_{\a \gd} = \nabla_\a R~,~~~~~~
	\nabla^\g W_{\a \b \g} = {\rm i} \,\nabla_{(\a }{}^\gd G_{\b) \gd}~.
	\eea
\end{subequations}
In this supergeometry integrals over the full superspace and its chiral subspace are related via:
\bea
\int \rd^{4|4}z \,E\, U 
= -\frac 14 \int \rd^{4|2}z_{+} \,
\,\cE\, (\bar{\nabla}^2 -4R) U 
~.
\eea

Constraints \eqref{SYMAlgebra} may be solved \cite{Wess:1977xn}
in terms of the complex unconstrained prepotential $\O = \O^I T^I$ and its Hermitian conjugate $\bar{\O} \equiv \O^\dagger$,
\begin{align}
	\label{2.4}
	\nabla_\a = \re^{-\O} \mathfrak{D}_\a \re^{\O} ~, \qquad \bar{\nabla}_\ad = \re^{\bar \O} \bar{\mathfrak D}_\ad \re^{-\bar{\O}}~, \qquad \re^{V} = \re^{\O} \re^{\bar{\O}}~,
\end{align}
where $\mathfrak{D}_A = (\mathfrak{D}_a, \mathfrak{D}_\a , \bar{\mathfrak D}^\ad )$ are the covariant derivatives of the GWZ superspace \cite{GWZ}, see e.g. \cite{Buchbinder:1998qv,Superspace} for a review.\footnote{Relation of the GWZ superspace geometry to other $\cN=1$ supergravity formulations in spelled out in \cite{KRT-M_N=1}. Our normalisation of the gauge prepotentials in \eqref{2.4} differs from \cite{Buchbinder:1998qv} and agrees with \cite{Superspace}.}
The superfields $\bar{\chi}_\ad $ and $\bar{W}_\ad $ in \eqref{N=1FT}
are Hermitian conjugates of $\chi_\a$ and $W_\a$, respectively, in the vector representation. The conjugation rules are somewhat different in the chiral representation \cite{Superspace}.

Making use of \eqref{2.2}, it follows 
that action \eqref{N=1FT} is invariant under the following gauge transformation
\begin{align}
	\label{1stGT}
	\d_{L} \chi_\a = - \frac \ri 4 ( \bar{\nabla}^2 - 4 R ) \nabla_\a L ~, \qquad L = \bar{L}~,
\end{align}
where the gauge parameter $L= L^I T^I$ is Hermitian but otherwise unconstrained. 
It follows that the algebra of gauge transformations is Abelian and closed, $\d_{L'} \d_{L} \chi_\a = 0$.
Actually, the super FT model is a reducible first-stage gauge theory, following the terminology of \cite{BV}.
Indeed, shifting the gauge parameter in \eqref{1stGT} as
\begin{subequations}
	\label{2ndGT}
\begin{align} 
L ~\to~L+\d_\s L~, \qquad 
	\d_{\s} L = \s + \bar{\s} ~, \quad \bar{\nabla}_\ad \s = 0~,
\end{align}
we obtain 
\begin{align}
	\label{2ndGT-b}
	\d_\s \d_L \chi_\a = \ri \big[ W_\a, \s\big]~,  \qquad 
	W_\a = - 2 \frac{ \d_{r} S_0}{\d \chi^\a}   
	~,
\end{align}
\end{subequations}
and therefore $\d_\s \d_L \chi_\a $ vanishes on the mass shell. 
An important implication of \eqref{2ndGT-b} is that the modified Faddeev-Popov quantisation procedures developed, e.g., in \cite{Siegel, Obukhov,BK88} cannot be directly applied to quantise the super FT model. 


\subsection{Classical equivalence}
\label{Section2.1}

The tensor multiplet is known to admit self-interactions of the form \cite{Siegel-tensor, LR83}
\bea
S =  \int \rd^{4|4} z \, E \, {\mathfrak F}  (G) ~, \qquad G= \hf ( \mathfrak{D}^\a \eta_\a  + \bar{\mathfrak D}_\ad \bar \eta^\ad) ~, \qquad 
\bar{\mathfrak D}_\bd \eta_\a =0~.
\label{SITM}
\eea
Here $\eta_\a$ denotes the chiral prepotential of a single tensor multiplet, to distinguish this case from the super FT model. The choice ${\mathfrak F} (G) = - \hf G^2$ corresponds to the free tensor multiplet, 
while ${\mathfrak F} (G) = -\m \, G \log (G/\m) $, with $\m$ a positive constant, 
describes the so-called improved tensor multiplet \cite{deWR}.
The model has a classically equivalent (or dual) formulation \cite{Siegel-tensor, LR83} realised in terms of a chiral scalar $\vf $ and its conjugate 
$\bar \vf$ with the action 
\bea
S =  \int \rd^{4|4} z \, E \, {\mathfrak U} (\vf + \bar \vf ) ~, \qquad \bar{\mathfrak D}_\bd \vf =0~,
\label{NLSM2}
\eea
where ${\mathfrak U} $ is the Legendre transform of ${\mathfrak F} $ \cite{LR83}, 
\bea
{\mathfrak U} = {\mathfrak F}(h) - h (\vf +\bar \vf) ~, \qquad {\mathfrak F}' (h) = \vf +\bar \vf~. 
\eea 
The target space of the supersymmetric $\s$-model \eqref{NLSM2} is a K\"ahler manifold with a $\sU(1)$ isometry group consisting of holomorphic transformations $\vf \to \vf + \ri r$, with $r \in {\mathbb R}$. 

On general grounds \cite{FradkinT}, the models \eqref{SITM} and \eqref{NLSM2} should remain equivalent in the quantum theory. Direct studies of their quantum equivalence in a supergravity background was carried out in \cite{GNSZ,BK88} for the case   ${\mathfrak F} (G) = - \hf G^2$. 

It was shown in \cite{CLL} that the super FT theory \eqref{N=1FT} is classically equivalent to a certain supersymmetric nonlinear sigma model. To see this, we first note that the dynamical equations for $\chi_\a$ and its conjugate imply that $V$ is pure gauge
\begin{align}
	\label{flatconnection}
	\frac{\d_l S_0[\chi,\bar{\chi},V]}{\d \chi^\a} = \hf W_\a = 0 \quad \implies \quad V =  \log ( \re^{\bar{\F}} \re^{\F} \big)~, \qquad \bar{\mathfrak D}_\ad \F = 0~.
\end{align}
As a result, the action \eqref{N=1FT} reduces to
\begin{align}
	\label{NLSM}
	S[\Phi,\bar{\Phi}] = \hf \text{tr} \int \rd^{4|4} z \, E \, \log^2(\re^{\bar{\F}} \re^{\F}) ~.
\end{align}
This is a supersymmetric nonlinear $\s$-model with K\"{a}hler potential 
\begin{align}
	\label{Kahler}
	K(\F,\bar{\F}) = \hf \tr \Big( \log^2(\re^{\bar{\F}} \re^{{\F}} )\Big)~.
\end{align}

One of the goals of this section, which we pursue below, is to prove that this equivalence extends to the quantum level. Specifically, our goal is to demonstrate that the partition functions for models \eqref{N=1FT} and \eqref{NLSM} differ by a topological invariant. This means that the quantum supercurrents in these theories coincide.

\subsection{Quantisation via Batalin-Vilkovisky formalism}

This subsection is devoted to quantising the super FT model \eqref{N=1FT}. 
The covariant quantisation of tensor multiplet models \eqref{SITM} in a supergravity background was carried out in 
 \cite{GNSZ,BK88} the modified Faddeev-Popov procedures.
As $\d_\s \d_L \chi_\a$ vanishes only on the equations of motion, see eq. \eqref{2ndGT-b},
these quantisation procedures are no longer directly applicable in the super FT case.
Thus, we will proceed by making use of the Batalin-Vilkovisky approach \cite{BV}.

As a first step, it is necessary to mathematically describe the structure of this model's gauge algebra. First, we recall that the generators of gauge transformations $R^i{}_{\underline{\a}}$ are defined via
\begin{align}
	\d_\xi \vf^i = R^i{}_{\underline{\a}} \xi^{\underline{\a}}~,
\end{align}
where $\vf = \{ \chi^{\a I}, \bar{\chi}_\ad^I, V^I \}$ and $\xi = \{ L^I \}$ denote the (classical) configuration space and gauge parameters, respectively. These generators can be read off from eq. \eqref{1stGT}:
\begin{subequations}
	\label{2.12}
	\begin{align}
		\d_L \chi^{\a I} (z) &= \int \rd^{4|4} z \, E \, {R}^{\a I \, J} (z,z') L^J(z')\quad \implies \quad {R}^{\a I \, J}(z,z') = - \frac{\ri}{4} (\bar{\nabla}^2 - 4 R) \nabla^\a \d^{8}(z,z') \d^{IJ}~, \\
		\d_L \bar{\chi}_\ad^{I} (z) &= \int \rd^{4|4} z \, E \, {R}_\ad^{I \, J} (z,z') L^J(z')\quad \implies \quad {R}_\ad^{I \, J}(z,z') = \frac{\ri}{4} (\nabla^2 - 4 \bar{R})\bar{\nabla}_\ad \d^{8}(z,z') \d^{IJ}~.
	\end{align}
\end{subequations}
We also recall that, in the present model, the gauge parameters $L^I$ possess their own gauge symmetry, see eq. \eqref{2ndGT}. This may be equivalently understood as the property that $R^i{}_{{\underline{\a}}}$ has nontrivial null eigenvectors $Z^{\underline{\a}}{}_{\underline{\a}_1}$ at a stationary point of the action:
\begin{align}
	\d_\z \xi^{\underline{\a}} = Z^{\underline{\a}}{}_{\underline{\a}_1} \z^{\underline{\a_1}} \quad \iff \quad R^i{}_{{\underline{\a}}} Z^{\underline{\a}}{}_{\underline{\a}_1} = 2 \frac{\d_r S_0}{\d \vf^j} K^{ji}_{\underline{\a}_1} (-1)^{\e_j} ~,
\end{align}
where the $K^{ji}_{\underline{\a}}$ term expresses the linear dependence of the generators off-shell. In the present case, the nontrivial null eigenvectors $Z^{\underline \a}{}_{\underline \a_1}$ of the generators are:
\begin{subequations}
	\label{2.14}
	\begin{align}
		\d_{\s} L^I(z) &= \int \rd^{4|2} z_+ \, \cE \, Z^{I \, J}(z,z') \s^J(z') + \int \rd^{4|2} z_- \, \bar{\cE} \, \bar{Z}^{I\,J}(z,z') \bar{\s}^J(z') ~, \\
		\implies& \quad Z^{I\,J}(z,z') = \d_{+}(z,z') \d^{IJ} ~, \qquad \bar{Z}^{I\,J}(z,z') = \d_{-}(z,z') \d^{IJ} ~.
	\end{align}
\end{subequations}
Finally, the $K^{ji}_{\underline \a_1}$ terms may be readily extracted from eq. \eqref{2ndGT}:
\begin{subequations}
	\label{2.15}
	\begin{align}
		K^{\a I \, \b J \, K}(z,z',z'') &= \ve^{\a \b} f^{IJK} \d_{+}(z,z') \d_+(z,z'') ~, \\
		\bar{K}_{\ad \, \bd}^{I \, J \, K}(z,z',z'') &= - \ve_{\ad \bd} f^{IJK} \d_{-}(z,z') \d_-(z,z'')~.
	\end{align}
\end{subequations}

Having described the gauge algebra structure of \eqref{N=1FT}, we are now equipped to quantise the theory. First, we enlarge the configuration space $\vf^i$ to the set of fields $\bm{\F}^{\underline{A}}$, which includes the appropriate ghosts. In addition, we associate with each field $\bm{\F}^{\underline{A}}$ an antifield $\bm{\F}^*_{\underline{A}}$ of opposite statistics; $\e(\bm{\F}^*_{\underline A}) = \e(\bm{\F}^{\underline A}) + 1$. The classical action $S_0[\vf]$ is then extended to a functional 
$S[\bm{\F},\bm{\F}^*]$ which is subject to the boundary condition $S[\bm{\F},\bm{\F}^*]\big|_{\bm{\F}^* = 0} = S_0[\vf]$ and is a proper solution of the master equation
\begin{align}
	\label{mastereq}
	\frac{\d_r S}{\d {\bm \F}^{\underline{A}}} \frac{\d_l S}{\d \bm{\F}_{\underline A}^*} = 0~.
\end{align}
Then, the in-out vacuum amplitude for $S_0[\vf]$ may be defined as
\begin{align}
	\label{PathIntegral}
	Z = \int \big[\mathscr{D} \bm{\F}^{\underline{A}} \big] \big[\mathscr{D}  \bm{\F}^*_{\underline{A}} \big] \, \mu \,  \d \Big[ \bm{\F}^*_{\underline{A}} - \frac{\d \Psi}{\d {\bm \F}^{\underline{A}}} \Big] \exp \Big( \ri S[\bm{\F},\bm{\F}^*] \Big)~,
\end{align}
where $\mu$ is the quantum integration measure and the fermionic functional $\Psi[\bm{\F}]$ is the gauge fermion, whose only requirement is that $Z$ possesses no residual gauge freedom.

For first-stage reducible theories a proper solution of the master equation \eqref{mastereq} always exists for the so-called ``minimal'' set of fields
\begin{align}
	\bm{\F}_\text{min} = \Big\{ \vf^i, C^{\underline{\a}} , \eta^{\underline{\a}_1}\Big\} ~, \qquad \e(C^{\underline{\a}}) = \e_{\underline{\a}} + 1~, \quad \e(\eta^{\underline{\a}_1}) = \e_{\underline{\a}_1}~,
\end{align}
which, in the case of an Abelian and closed gauge algebra, takes the following form
\begin{align}
	\label{2.21}
	S[\bm{\F}_\text{min},{\bm{\F}}^*_\text{min}] = S_0[\vf] + \vf^*_i R^i{}_{\underline{\a}} C^{\underline{\a}} + C^*_{\underline{\a}} Z^{\underline{\a}}{}_{\underline{\a}_1}\eta^{\underline{\a}_1} + \vf^*_i \vf^*_j K^{ji}_{\underline{\a}_1} \eta^{\underline{\a}_1}~.
\end{align}
Making use of equations \eqref{2.12}, \eqref{2.14} and \eqref{2.15}, we find $S[\bm{\F}_\text{min},{\bm{\F}}^*_\text{min}]$ to be:
\begin{align}
	\label{MinimalSoln}
	S[\bm{\F}_\text{min},{\bm{\F}}^*_\text{min}] &= S_0[\chi,\bar{\chi},V] + \text{tr} \int \rd^{4|4} z \, E \, \Big \{ \ri \big( \nabla^\a \chi_\a^* - \bar{\nabla}_\ad \bar{\chi}^{* \ad}\big) C + C^* (\eta + \bar{\eta})\Big \} \non \\ 
	& \phantom{=} + \ri \, \text{tr} \int \rd^{4|2} z_+ \, \cE \,  \big[ \chi^{* \a} , \chi_\a^* \big] \eta  - \ri \, \text{tr} \int \rd^{4|2} z_- \, \bar{\cE} \, \big[ \bar{\chi}^{*}_\ad , \bar{\chi}^{\ad *} \big]  \bar{\eta} ~,
\end{align}
where the ghost superfield $\eta$ is covariantly chiral, $\bar{\nabla}_\ad \eta = 0$.

Finally, it is necessary to construct a useful gauge fermion $\Psi(\bm{\F})$ to eliminate the antifields present in \eqref{MinimalSoln} in accordance with \eqref{PathIntegral}. To this end, we extend the set of fields from $\bm{\F}_\text{min}$ to $\bm{\F}$ by appending the following pairs of Lie algebra-valued superfields:
\begin{subequations}
	\label{nonminimal}
	\begin{align}
		(i):& \qquad (\tilde{C},\pi)~, \quad \e(\tilde{C}) = \e(\pi) + 1 = 1~, \label{nonminimal-a} \\
		(ii):& \qquad (\bm{\eta},\bm{\pi})~, \quad \bar{\nabla}_\ad \bm{\eta} = 0 ~, \quad \bar{\nabla}_\ad{\bm{\pi}} = 0~, \qquad \e(\bm{\eta}) = \e(\bm{\pi}) - 1 = 0~, \label{nonminimal-b} \\
		(iii):& \qquad ({\eta}',{\pi}')~, \quad \bar{\nabla}_\ad \eta' = 0 ~, \quad \bar{\nabla}_\ad \pi' = 0~, \qquad \e({\eta}') = \e({\pi}') - 1 = 0~, \label{nonminimal-c} \\
		(iv):& \qquad (\L,\Pi)~, \quad \bar{\nabla}_\ad \L = 0 ~, \quad \bar{\nabla}_\ad \Pi = 0~, \qquad \e(\L) = \e(\Pi) + 1 = 1~, \label{nonminimal-d} \\
		(v):& \qquad (\L',\Pi')~, \quad \bar{\nabla}_\ad \L' = 0 ~, \quad \bar{\nabla}_\ad \Pi' = 0~, \qquad \e(\L') = \e(\Pi') + 1 = 1~. \label{nonminimal-e}
	\end{align}
\end{subequations}
These will be used to obtain a new solution of master equation \eqref{mastereq} that is more useful for the elimination of antifields. It is then easily seen that the following functional
\begin{align}
	\label{extendedaction}
	S[\bm{\F},\bm{\F}^*] &= S[\bm{\F}_\text{min},{\bm{\F}}^*_\text{min}]
	+ \text{tr} \int \rd^{4|4} z \, E \, \Big \{ \tilde{C}^* \pi + \hf \Big [(\bm{\eta}^* + \bar{\bm{\pi}})^2 + (\bar{\bm{\eta}}^* + \bm{\pi})^2 + (\eta'^* + \bar{\pi}')^2  \non \\
	& \qquad  + (\bar{\eta}'^* + \pi')^2 + (\L^* + \bar{\Pi})^2 + (\bar{\L}^* + \Pi)^2 + (\L'^* + \bar{\Pi}')^2 + (\bar{\L}^* + \Pi')^2 \Big] \Big \}
\end{align}
is also a solution to the master equation \eqref{mastereq}. It should be noted that this functional satisfies
\begin{align}
	\D S[\bm{\F} , \bm{\F}^*] = 0 ~, \qquad \D := \frac{\d_r}{\d {\bm \F}^{\underline{A}}} \frac{\d_l}{\d {\bm \F}^*_{\underline{A}}} ~,
\end{align}
therefore the quantum measure in \eqref{PathIntegral} may be set to unity, $\mu = 1$, in accordance with \cite{BaKa}.
Choosing the gauge fermion to be
\begin{align}
	\Psi(\bm{\F}) &= \text{tr} \int \rd^{4|4} z \, E \, \Big \{ \frac \ri 2 \tilde{C} \big(\nabla^\a \chi_\a - \bar{\nabla}_\ad \bar{\chi}^\ad \big) + C(\bm{\eta} + \bar{\bm{\eta}}) + \tilde{C} (\eta' + \bar{\eta}') - \frac{1}{2m} \tilde{C} \pi \Big \} \non \\
	&\phantom{=} \quad + \frac 1 2 \text{tr} \int \rd^{4|2} z_+ \, \cE \, \Big \{ -\frac{1}{n} \big({\bm{\eta}}\pi' + \bm{\pi} \eta'\big) + \L \Pi + \L' \Pi' \Big \} \non \\
	&\phantom{=} \quad + \frac 1 2 \text{tr} \int \rd^{4|2} z_- \, \bar{\cE} \, \Big \{ -\frac{1}{n} \big(\bar{\bm{\eta}}\bar{\pi}' + \bar{\bm \pi} \bar{\eta}'\big) + \bar{\L} \bar{\Pi} + \bar{\L}' \bar{\Pi}' \Big \}~,
\end{align}
and eliminating antifields in eq. \eqref{extendedaction} leads to
\begin{align}
	\label{2.25}
	S[\bm{\F},{\bm{\F}}^*] &= S_0[\vf] + \text{tr} \int \rd^{4|4} z \, E \, \Big \{ \frac 1 8 \tilde{C} \big(\nabla^\a (\bar{\nabla}^2 - 4 R) \nabla_\a + \bar{\nabla}_\ad (\nabla^2 - 4 \bar{R}) \bar{\nabla}^\ad \big) C + \hf(\bar{\eta} + \bm{\eta})^2 \non \\
	& \phantom{=} \quad + \hf(\bar{\bm{\eta}} + \eta)^2 - \hf (\bar{\Pi} + \Pi)^2 - \hf (\bar{\Pi}' + \Pi')^2 + \Big( \frac{\ri}{2} \big(\nabla^\a \chi_\a - \bar{\nabla}_\ad \bar{\chi}^\ad \big) + (\eta' + \bar{\eta}') - \frac{1}{2m} \pi \Big) \pi  \non \\
	& \phantom{=} \quad + \hf \Big( - \frac{1}{2n} \pi' - \frac 1 4 (\bar{\nabla}^2 - 4 R) C + \bar{\bm \pi}\Big)^2 + \hf \Big( - \frac{1}{2n} \bar{\pi}' - \frac 1 4 (\nabla^2 - 4 \bar{R}) C + \bm{\pi} \Big)^2 \non \\
	& \phantom{=} \quad  + \hf \Big( \frac{1}{2n} \bm{\pi} - \frac 1 4 (\bar{\nabla}^2 - 4 R) \tilde{C} + \bar{\pi}'\Big)^2  + \hf\Big( \frac{1}{2n} \bar{\bm \pi} - \frac 1 4 (\nabla^2 - 4 \bar{R}) \tilde{C} + \pi' \Big)^2   \Big \} \non \\
	&\phantom{=} \quad - \frac \ri {64} \, \text{tr} \int \rd^{4|2} z_+ \, \cE \, \big[ (\bar{\nabla}^2 - 4 R) \nabla^\a \tilde{C} , (\bar{\nabla}^2 - 4 R) \nabla_\a \tilde{C} \big] \eta  \non \\
	&\phantom{=} \quad + \frac \ri {64} \text{tr} \int \rd^{4|2} z_- \, \bar{\cE} \, \big[ (\nabla^2 - 4 \bar{R}) \bar{\nabla}_\ad \tilde{C}, (\nabla^2 - 4 \bar{R}) \bar{\nabla}^\ad \tilde{C} \big] \bar{\eta} ~.
\end{align}
Now, performing the path integral over $\pi$, $\bm{\pi}$, $\pi'$, $\Pi$, $\Pi'$ and $\bar{\bm{\pi}}$, $\bar{\pi}'$, $\bar{\Pi}$, $\bar{\Pi}'$, we obtain a new representation for the in-out vacuum amplitude \eqref{PathIntegral}
\begin{align}
	Z &= \int \big[\mathscr{D} \bm{\vf} \big] \exp \Big( \ri S[\bm{\vf}] \Big)~, \qquad \bm{\vf} = \Big \{ \chi^\a , \bar{\chi}_\ad, V, C, \tilde{C}, \eta, \bar{\eta}, \bm{\eta}, \bar{\bm{\eta}}, \eta', \bar{\eta}' \Big \} ~, \non \\
	S[\bm{\vf}] &= S_0[\vf] + \text{tr} \int \rd^{4|4} z \, E \, \Big \{ \frac 1 8 \tilde{C} \Big(\nabla^\a (\bar{\nabla}^2 - 4 R) \nabla_\a + \bar{\nabla}_\ad (\nabla^2 - 4 \bar{R}) \bar{\nabla}^\ad \non \\
	& \phantom{=} \quad + \frac{n}{2} \big [ ({\nabla}^2 - 4 \bar{R}) + (\bar{\nabla}^2 - 4 R) \big ]^2 \Big) C +\hf(\bar{\eta} + \bm{\eta})^2 + \hf (\bar{\bm{\eta}} + \eta)^2 \non \\
	& \phantom{=} \quad + \frac{m}{2} (\bar{\eta}' + \eta')^2 - \frac{m}{8} \big( \nabla^\a \chi_\a - \bar{\nabla}_\ad \bar{\chi}^\ad \big)^2 \Big \} \non \\
	&\phantom{=} \quad - \frac \ri {64} \, \text{tr} \int \rd^{4|2} z_+ \, \cE \, \big[ (\bar{\nabla}^2 - 4 R) \nabla^\a \tilde{C} , (\bar{\nabla}^2 - 4 R) \nabla_\a \tilde{C} \big] \eta  \non \\
	&\phantom{=} \quad + \frac \ri {64} \text{tr} \int \rd^{4|2} z_- \, \bar{\cE} \, \big[ (\nabla^2 - 4 \bar{R}) \bar{\nabla}_\ad \tilde{C}, (\nabla^2 - 4 \bar{R}) \bar{\nabla}^\ad \tilde{C} \big] \bar{\eta} ~.
\end{align}

To conclude this subsection, some nontrivial aspects of the above analysis should be clarified. Specifically, in order to eliminate antifields in action \eqref{MinimalSoln}, we extended $\bm{\F}_\text{min}$ by five pairs of superfields \eqref{nonminimal}. This is in contrast to the usual scheme of \cite{BV}; only the three pairs \eqref{nonminimal-a} - \eqref{nonminimal-c} should be necessary. The novelty in our construction originates from the property that pairs \eqref{nonminimal-b} and \eqref{nonminimal-c} necessarily appear in \eqref{2.25} with their kinetic operators,\footnote{This feature was first noticed in \cite{Grisaru:1997hf}, in the context of quantisation of the complex linear superfield.} which provide nontrivial contributions to $Z$. To counteract this, we introduced superfields of opposite statistics \eqref{nonminimal-d} and \eqref{nonminimal-e}, which negate these superficial degrees of freedom.


\subsection{Quantum equivalence}
\label{section2.3}

Having quantised the super FT model \eqref{N=1FT} above, we now specialise our analysis to the extension of classical equivalence demonstrated in section \ref{Section2.1} to the quantum level. Our analysis will be similar to that given in \cite{BK87} in the non-supersymmetric case. 

Our starting point will be the functional \eqref{2.25}, which can be simplified by: (i) noting that the three-ghost terms do not contribute to the path integral;\footnote{This is clear if one performs the path integral over $\bm{\eta}$ and $\bar{\bm{\eta}}$.} (ii) taking the $m \rightarrow \infty$ limit and setting $n=-2$; and (iii) performing the path integral over $\chi^\a$, $\eta'$, $\bm{\pi}$, $\pi'$, $\Pi$, $\Pi'$ and $\bar \chi_\ad$, $\bar \eta'$, $\bar{\bm{\pi}}$, $\bar \pi'$, $\bar \Pi$, $\bar \Pi'$. 
The resulting in-out vacuum amplitude is:
\begin{align}
	\label{2.26}
	Z &= 
	\int \big[\mathscr{D} V \big] \big[\mathscr{D} \tilde{C} \big] \big[\mathscr{D} C \big] \big[\mathscr{D} \bar{\eta} \big] \big[\mathscr{D} \eta \big] \big[\mathscr{D} \bar{\bm{\eta}} \big] \big[\mathscr{D} \bm{\eta} \big] \big[\mathscr{D} \pi \big] \d \Big((\nabla^2 - 4 \bar{R}) \pi \Big) \d\Big((\bar{\nabla}^2 - 4 R)\pi\Big) \non \\
	&\qquad \qquad \times \d \Big(W_\a + \frac{\ri}{4} (\bar{\nabla}^2 - 4 {R} ) \nabla_\a \pi \Big) \d \Big(\bar{W}_\ad - \frac{\ri}{4} (\nabla^2 - 4 \bar{R}) \bar{\nabla}_\ad \pi \Big) \text{Det}(\mathcal{H}) \non \\
	& \qquad \qquad \times \exp \Big ( \ri \,\text{tr} \int \rd^{4|4} z \, E \, \Big \{ V^2 - 2 \tilde{C} \Box_V C + \hf ({\bm{\eta}} + \bar{\eta})^2 + \hf ({\eta} + \bar{\bm{\eta}})^2\Big \}\Big )~,
\end{align}
where we have introduced the operators 
\begin{subequations}
\begin{align}
	\mathcal{H} &= 
	\begin{pmatrix}
		R & -\frac 1 4 (\bar{\nabla}^2 - 4 R) \\ 
		-\frac 1 4 ({\nabla}^2 - 4 \bar{R}) & 
		\bar{R}
	\end{pmatrix} ~, \\
	\Box_V & = - \frac{1}{16} \Big( \nabla^\a (\bar{\nabla}^2 - 4 R) \nabla_\a +  \bar \nabla_\ad  ({\nabla}^2 - 4 \bar{R}) \bar \nabla^\ad 
	-\big [ ({\nabla}^2 - 4 \bar{R}) + (\bar{\nabla}^2 - 4 R) \big ]^2 \Big)\non \\
	&= \nabla^a \nabla_a - \frac 14 G^{\aa} [\nabla_\a , \nabla_\ad] - \frac14 \bar{\nabla}_\ad \bar{R} \bar{\nabla}^\ad - \frac14 \nabla^\a R \nabla_\a - \frac14 \bar{\nabla}^2 \bar{R} - \frac14 \nabla^2 R  \non \\
	&\phantom{=} + 2 R \bar{R} - \frac 14 R (\bar{\nabla}^2 - 4 R) - \frac 14 \bar{R} ({\nabla}^2 - 4 \bar{R}) - \frac 14 W^\a \nabla_\a + \frac 14 \bar{W}_\ad \bar{\nabla}^\ad ~.
\end{align}
\end{subequations}
The operator $\mathcal{H} $ acts on the space of covariantly chiral-antichiral pairs in the adjoint representation of $G$,
while $\Box_V$ is defined to act on the space of unconstrained scalar superfields in the adjoint representation of $G$. The operators $\mathcal{H} $  and $\Box_V$ are gauge-covariant extensions of the operators introduced in \cite{BK86}.

To evaluate the $\d$-functions in \eqref{2.26}, we introduce a background quantum splitting \cite{Grisaru:1979wc} of the prepotentials in  \eqref{2.4}
\bea
\re^\O = \re^\o \, \re^{\O_Q}~, \qquad \re^{V} = \re^\o \,\re^{V_Q} \, \re^{\bar \o}~.
\label{BQsplitting}
\eea
Here $\o$ denotes the background prepotential and $V_Q$ the quantum gauge superfield. We denote by $\cD_A = (\cD_a, \cD_\a , \bar \cD^\ad )$ the background covariant derivatives. We will use the vector representation for the background covariant derivatives, and the chiral quantum representation for the original operators 
$\nabla_A$,
\bea
\nabla_\a = \re^{-V_Q} \cD_\a \re^{V_Q}~, \qquad \bar \nabla_\ad = \bar \cD_\ad~.\eea
The field strength $W_\a$ takes the form 
\bea
W_\a = \re^{\bar \O_Q} \left( -\frac 14 (\bar \cD^2 - 4 {R}) \Big( \re^{-V_Q} \cD_\a \re^{V_Q} \Big) + w_\a\right)\re^{-\bar \O_Q} ~,
\eea
where $w_\a$ is the covariantly chiral background field strength, $\bar \cD_\bd w_\a=0$. 
 Since we are interested in evaluating the $\d$-functions in \eqref{2.26}, 
we decompose $W_\a$ to first order in the quantum superfields,
\bea
W_\a \approx - \frac 14  (\bar \cD^2 - 4 {R}) \cD_\a V_Q +w_\a + \big[ \bar \O_Q, w_\a\big]~.
\eea
The $\d$-functions in \eqref{2.26} tell us that the background connection must be chosen to be flat, 
\bea
w_\a=0 \quad \implies \quad \re^{\bar \o} = \re^\F~, \qquad {\bar{\mathfrak{D}}}_\ad \F =0~.
\label{FlatConn}
\eea
In evaluating $\d(W_\a + \frac{\ri}{4} (\bar{\nabla}^2 - 4 {R}) \nabla_\a \pi) $, it suffices to keep only the terms of first order in the quantum prepotentials and $\p$,
\bea
\d \left(W_\a + \frac{\ri}{4} (\bar{\nabla}^2 - 4 R) \nabla_\a \pi\right) 
= \d \left( - \frac 14  (\bar{\cD}^2 - 4 {R}) \cD_\a V_Q +  \frac{\ri}{4}  (\bar{\cD}^2 - 4 {R}) \cD_\a \p\right)~.
\eea
Now we consider the following change of variables:
\begin{subequations}
	\begin{align}
		V_Q &= \r + \bar{\r} + \hf \big( \cD^\a \l_\a + \bar{\cD}_\ad \bar{\l}^\ad \big)~, \qquad \bar{\cD}_\ad \r = 0~, \quad \cDB_\ad \l_\a = 0~, \\
		\pi &= \g + \bar{\g} + \frac{\ri}{2} \big( \cD^\a \l_\a - \bar{\cD}_\ad \bar{\l}^\ad \big)~, \qquad \bar{\cD}_\ad \g = 0~.
	\end{align}
\end{subequations}
Such a change of variables has a non-trivial Jacobian $\mathfrak{J} $.
It can be computed by applying this change of variables in the right-hand side of 
\begin{align}
	1 = \int \big[\mathscr{D} V_Q \big] \big[\mathscr{D} \p \big] \exp \Big ( \frac{\ri}{2} \text{tr} \int \, \rd^{4|4}z \, E \, \Big \{ {V_Q}^2 -\p^2 \Big \}  \Big )
\end{align}
to result with
\begin{align}
	\label{Jacobian}
	\mathfrak{J} = \text{Det}(\widetilde{\cH}) \text{Det}_+^{-1/2}(\widetilde{\mathscr{H}}_c) 
	\text{Det}_-^{-1/2}(\widetilde{\mathscr{H}}_a)~,
\end{align}
where $\widetilde{\mathscr{H}}_c$ ($\widetilde{\mathscr{H}}_a$) is a d'Alembertian on the space of covariantly chiral (antichiral) spinor superfields
\begin{align}
	\widetilde{\mathscr{H}}_c = \frac{1}{16} (\bar{\cD}^2 - 4 {R}) (\cD^2 - 6 \bar{R}) ~, \qquad 
	\widetilde{\mathscr{H}}_a = \frac{1}{16} (\cD^2 - 4 \bar{{R}}) (\bar{\cD}^2 - 6 {{R}})~.
\end{align}
Here and below, a tilde over operators, such as $\widetilde{\cH} $,
  indicates that these operators are constructed using the flat-connection covariant derivatives $\cD_A$.
With this change of variables, we find that eq. \eqref{2.26} simplifies to
\begin{align}
	\label{2.43}
	Z &=  \int \big[\mathscr{D} \bar{\F} \big] \big[\mathscr{D} {\F} \big] \text{Det}_+^{1/2}(\widetilde{\mathscr{H}}_c) \text{Det}_-^{1/2}(\widetilde{\mathscr{H}}_a) \text{Det}(\widetilde{{\square}}_V)
	 \text{Det}^{-2}(\widetilde{\mathcal{H}}) \exp \Big ( \ri S[\F,\bar{\F}] \Big ) ~.
\end{align}
The quantum gauge field is now equal to zero, $V_Q =0$, and the relations \eqref{BQsplitting} and \eqref{FlatConn}
tell us that 
\begin{align}
	V = \log ( \re^{\bar{\F}} \re^{\F} \big)~, \qquad \bar{\mathfrak D}_\ad \F = 0~.
\end{align}
Now, the analysis in \cite{BK88} (see also \cite{Buchbinder:1998qv} for a review)
implies that 
\begin{align}
	\label{DetIdentity}
	\text{Det}_+^{1/2}(\widetilde{\mathscr{H}}_c) \text{Det}_-^{1/2}(\widetilde{\mathscr{H}}_a) 
	\text{Det}(\widetilde{\square}_V) \text{Det}^{-2}(\widetilde{\mathcal{H}}) ~,
\end{align}
is a topological invariant, that is it does not depend on the supergravity prepotentials $H^{m}$, $\vf$ and $\bar \vf$, where $H^m$ is the gravitational superfield \cite{OS,Siegel78} and 
 $\vf$ the so-called chiral compensator \cite{Siegel78} (see \cite{Buchbinder:1998qv, Superspace} for a review).
Therefore, the in-out vacuum amplitude \eqref{2.43} reduces to that for the supersymmetric nonlinear $\sigma$-model \eqref{NLSM}
\begin{align}
	Z &=  \int \big[\mathscr{D} \bar{\F} \big] \big[\mathscr{D} {\F} \big] \exp \Big ( \ri S[\F,\bar{\F}] \Big ) ~.
\end{align}
 
 Thus we have demonstrated that the partition functions of the super FT model  \eqref{N=1FT} and 
 the supersymmetric nonlinear $\sigma$-model \eqref{NLSM} coincide, which implies the equality of the ``on-shell'' effective actions in these theories\cite{FradkinT}.
This establishes the quantum equivalence of the two theories  in the sense that the quantum supercurrents of the two theories coincide.  
One may also add a source term for $V$ to \eqref{N=1FT} and repeat the analysis to obtain formal equality of their partition functions. This is an extension of the results obtained in \cite{BK88, GNSZ}.



\section{Quantisation of $\cN=2$ tensor multiplet models}
\label{section3}

The $\cN=2$ tensor multiplet \cite{Wess} can be described by
its gauge-invariant field strength $G^{ij}$  which is defined to be a  real ${\sSU}(2)$ triplet (that is, 
$G^{ij}=G^{ji}$ and ${\bar G}_{ij}:=\overline{G^{ij}} = G_{ij}$)
subject to the covariant constraints  \cite{BS,SSW}
\bea
D^{(i}_\a G^{jk)} =  {\bar D}^{(i}_\ad G^{jk)} = 0~,
\label{analyticity1}
\eea
where $D_A = (\pa_a, D_\a^i , \bar D^{\ad}_i) $ are covariant derivatives of $\cN=2$ Minkowski superspace.
These constraints are solved in terms of a chiral
prepotential $\Psi$ \cite{HST,GS82,Siegel83,Muller86} as follows:
\begin{align}
\label{eq_Gprepotential}
G^{ij} = \frac{1}{8}D^{ij} \Psi
+\frac{1}{8}D^{ij} {\bar \Psi}~, \qquad
{\bar D}^i_\ad \J=0~,
\end{align}
where we have used the following notation:
$D^{ij}:= D^{\a(i} D_\a^{j)}$ and ${\bar D}^{ij} := \bar{D}_\ad^{(i} \bar{D}^{\ad j)}$.
The chiral prepotential $\J$ is defined modulo gauge transformations of the form 
\bea
\label{3.3}
\J \to  \J + {\rm i}  \L ~, \qquad
 {\bar D}_\ad^i \L =0~, \qquad 
D^{ij} \L 
= {\bar D}^{ij} {\bar \L}~,
\label{gauge-tr}
\eea
which leave $G^{ij}$ invariant. Here the gauge parameter $\L$ is a reduced chiral superfield, that is it satisfies the same constraints as the chiral field strength of the Abelian vector multiplet \cite{GSW}.


\subsection{Is there an $\cN=2$ supersymmetric FT model?}

The algebra of $\cN=2$ gauge-covariant derivatives $\nabla_A = (\nabla_a,\nabla_\a^i,\bar{\nabla}^\ad_i)$
is \cite{GSW}
\begin{subequations}
\bea
& \{ \nabla^i_\alpha,\bar{\nabla}_{{\dot\alpha j}} \}= -2{\rm i}\delta^i_j
\nabla_{\alpha{\dot\alpha}} ~, \\
&\{ \nabla^i_\alpha,\nabla^j_\beta\}={\rm i}\varepsilon_{\alpha\beta}
\varepsilon^{ij}{\bar W}~,\qquad \{{\bar \nabla}_{{\dot\alpha}i},
{\bar \nabla}_{{\dot\beta}j}\}={\rm i}\varepsilon_{{\dot\alpha}{\dot\beta}}
\varepsilon_{ij}W~,  \label{anticomm}\\
&{[}\nabla_{\alpha{\dot\alpha}},\nabla^j_\beta]=\hf \varepsilon_
{\alpha\beta}{\bar \nabla}^i_{\dot\alpha}{\bar W} ~,\qquad
[\nabla_{\alpha{\dot\alpha}},{\bar \nabla }_{{\dot\beta}i}]=\hf
\varepsilon_{{\dot\alpha}{\dot\beta}} \nabla_{\alpha i}W  ~,
\eea
\end{subequations}
where $W$ is a covariantly chiral superfield strength, $\bar \nabla^\ad_i W =0$, satisfying the Bianchi identity 
$\nabla^{ij} W = \bar \nabla^{ij} \bar W$. It follows from \eqref{anticomm} that a covariantly chiral matter multiplet $\f$, transforming in some representation of the gauge group $G$, does not exist,
\bea
\bar \nabla^\ad_i \f = 0 \quad \implies \quad W \f =0~.
\eea
This means that there is no obvious $\cN=2$ generalisation of the first two terms in \eqref{N=1FT}.
We therefore restrict our attention to Abelian tensor multiplets.


\subsection{The free tensor multiplet model}

The most general self-couplings of $\cN=2$ tensor multiplets derived in \cite{LR83} possess a simple manifestly $\cN=2$ supersymmetric formulation within the so-called projective superspace approach \cite{KLR, LR-projective}. From the point of view of the quantum theory, the harmonic-superspace 
\cite{GIKOS, GIOS} is advantageous. 
The two approaches are often complementary \cite{K98,K2010}.

Here we will use the harmonic-superspace formulation for the tensor multiplet \cite{GIO1,GIOS}.
A free $\cN=2$ tensor multiplet is described by the action 
\bea
S [\J, \bar \J]=  \hf 
\int {\rm d}\zeta^{(-4)}\,(G^{++})^2~,
\label{massless-tensor}
\eea
where 
the field strength \eqref{eq_Gprepotential} now takes the form 
\begin{subequations}
\bea
G^{++} (z,u) =G^{ij}(z) \,u^+_i u^+_j
= \frac{1}{8} (D^+)^2 \J (z)
+\frac{1}{8} ({\bar D}^+)^2 {\bar \J} (z) ~,
\label{3.7a}
 \eea
and the constraints \eqref{analyticity1} turn into
\bea
D^+_\a G^{++} =0~,
\label{3.7b}
\quad \bar D^+_\ad G^{++} =0~.
\eea
It also follows from \eqref{3.7a} that 
\bea
D^{++} G^{++}=0~.
\label{3.7c}
\eea
\end{subequations}

The action \eqref{massless-tensor} can be written in a first-order form 
\bea
S [\J, \bar \J, V^{++}]&=&  
\int {\rm d}\zeta^{(-4)}\,\left\{ G^{++}V^{++} - \hf (V^{++})^2\right\}~,
\eea
where $V^{++}$ is a real analytic superfield of $\sU(1)$ charge $+2$.
The first term in this action can be rewritten as a chiral integral
\bea
\int {\rm d}\zeta^{(-4)}\,
G^{++}V^{++} = 
\hf \left\{  \int {\rm d}^4 x \rd^4\q  \, \J W
+{\rm c.c.} \right\} ~.
\eea
Here 
 $W(z)$ is the (harmonic independent) 
chiral field strength of the $\cN=2$  vector multiplet
\cite{GSW}, 
\be
{\bar D}^i_\ad W =0~, \qquad 
D^{ij} W 
= {\bar D}^{ij} {\bar W}~, 
\ee
which is expressed via 
the analytic prepotential $V^{++} $
as follows \cite{GIKOS,GIOS,Z}:
\bea
W(z)= {1\over 4} \int {\rm d}u \, 
({\bar D}^-)^2 \,V^{++}(z,u)
~.
\label{n=2vmfs} 
\eea

The gauge freedom \eqref{gauge-tr} may be fixed by imposing the gauge condition 
$H^{++}=0$, where 
\bea
H^{++} =  \frac{\ri}{8} (D^+)^2 \J
-\frac{\ri}{8} ({\bar D}^+)^2\, {\bar \J}  ~.
\label{gaugecondition}
\eea
Choosing a gauge-fixing term 
\bea
S_{\rm gf}   [\J, \bar \J] = \hf \int {\rm d}\zeta^{(-4)}\,(H^{++})^2~,
\eea
we obtain 
\bea 
S [\J, \bar \J] + S_{\rm gf}   [\J, \bar \J]= \hf \int \rd^4 x \rd^4 \q \rd^4 \bar \q \, 
\bar \J \J \equiv \hf \int \rd^{4|8} z\, \bar \J \J~.
\label{GFaction}
\eea
Here we have used the integration rule 
\bea
 \int \rd^4 x \rd^4 \q \rd^4 \bar \q \, \mathscr{L} = 
  \int {\rm d}\zeta^{(-4)}\, (D^+)^4 \mathscr{L}~, 
\eea
where the fourth-order operator $(D^+)^4$ is defined in \eqref{A.5}. Gauge-fixed action \eqref{GFaction} is a higher-derivative model. 

The tensor multiplet is known to be dual to a real hypermultiplet, see e.g. \cite{LR83}. Within the harmonic-superspace approach, the tensor multiplet is dual to the so-called $\o$ hypermultiplet, which was introduced in \cite{GIKOS} and is described by the action 
\bea
\label{3.16}
S_\o = -\hf \int {\rm d}\zeta^{(-4)}\,(D^{++} \o)^2~, \qquad D^+_\a \o =0~,
\quad \bar D^+_\ad \o =0~,
\eea
where $\o$ is subject to be real (with respect to the analyticity-preserving conjugation). The duality is manifested by considering the first-order action \cite{GIO1}
\bea
S = \int {\rm d}\zeta^{(-4)}\,\left\{ \hf (U^{++})^2 + U^{++} D^{++} \o\right\} ~,
\qquad D^+_\a {U}^{++} =0~,
\quad \bar D^+_\ad {U}^{++} =0~,
\eea
with $U$ a real analytic superfield.


\subsection{Quantisation of the free tensor multiplet}
\label{section3.3}

The tensor multiplet model \eqref{massless-tensor} is a gauge theory with linearly dependent generators. The gauge parameter $\L$ in \eqref{gauge-tr} is a reduced chiral superfield which can be expressed in terms of a real analytic superfield 
$\mathfrak{U}^{++}$ of $\sU(1)$ charge $+2$,
\bea
\L[{\mathfrak U}^{++} ]= \frac{1}{4} \int {\rm d}u \, 
({\bar D}^-)^2 {\mathfrak U}^{++}~, \qquad D^+_\a {\mathfrak U}^{++} =0~,
\quad \bar D^+_\ad {\mathfrak U}^{++} =0~.
\label{reduced}
\eea
The chiral scalar $\L$ does not change upon the replacement  
\bea
{\mathfrak U}^{++} \to {\mathfrak U}^{++} + D^{++} \s ~, \qquad D^+_\a \s =0~,
\quad \bar D^+_\ad \s =0~.
\label{VM-gauge-tr}
\eea
This means that \eqref{massless-tensor} is a first-stage reducible theory, and therefore it cannot be quantised using the Faddeev-Popov procedure \cite{Faddeev:1967fc}.
In what follows, we will need another representation for the reduced chiral scalar
\eqref{reduced}, see e.g. \cite{GIOS}. It is
\bea
 \L[{\mathfrak U}^{++} ]= \frac{1}{4} 
({\bar D}^+)^2 {\mathfrak U}^{--}(z,u)~, \qquad 
{\mathfrak U}^{--} (z,u) = \int {\rm d} u' \, 
\frac{ {\mathfrak U}^{++}(z, u')}
{(u^+u'^+)^2} ~.
\label{V--}
 \eea

Several consistent quantisation procedures have been developed to quantise reducible Abelian  gauge theories such as gauge $p$-forms \cite{Schwarz1,Schwarz2, Siegel, Thierry-Mieg:1980ihu, Obukhov, BK88},   
 including the formulations of \cite{Siegel,BK88} which have been applied in the $\cN=1$ supersymmetric case.  
 These quantisation schemes are much easier to deal with than the 
 Batalin-Vilkovisky formalism \cite{BV}. Here we will make use of the modified Faddeev-Popov procedure described in \cite{BK88} to quantise the tensor multiplet model  \eqref{massless-tensor}.\footnote{Quantisation of this model within the BV formalism is described in appendix \ref{appendixB}.}

Our first step is to define a generalised delta-function $\hat{\d} [H^{++}]$, with $H^{++}$ the gauge fixing function 
\eqref{gaugecondition}. The latter is a real analytic superfield obeying the constraint $D^{++} H^{++}=0$, and therefore
the naive delta-function 
\begin{align}
\d \big[H^{++}  \big]= \int \big[\mathscr{D} {\mathfrak V}^{++}\big] \exp \left\{ \ri \int {\rm d}\zeta^{(-4)}\,{\mathfrak V}^{++} H^{++} \right \}~, \qquad D^+_\a {\mathfrak V}^{++} =0~,
\quad \bar D^+_\ad {\mathfrak V}^{++} =0~,
\end{align}
is ill-defined since the exponential  is invariant under the gauge transformations
\bea
{\mathfrak V}^{++} \to {\mathfrak V}^{++} + D^{++} \s ~, \qquad D^+_\a \s =0~,
\quad \bar D^+_\ad \s =0~.
\label{gauge-sigma}
\eea
This is the gauge freedom for the Abelian vector multiplet, and it can be fixed by applying the Faddeev-Popov procedure to result with 
\begin{subequations}\label{gen-delta} 
\bea
\hat{\d } \big[H^{++}  \big]&=& \int \big[\mathscr{D} {\mathfrak V}^{++}\big] \,
 \d\big[ D^{++} {\mathfrak V}^{++} \big] \D_{\rm VM}
 \exp \left\{ \ri \int {\rm d}\zeta^{(-4)}\, {\mathfrak V}^{++} H^{++} \right\} ~,
\label{gen-delta.a}  \\
\D_{\rm VM}&=& \int \big[\mathscr{D} \tilde{C} \big] \big[\mathscr{D} C \big]  
\exp \left\{ \ri \int {\rm d}\zeta^{(-4)}\,\tilde{C} (D^{++})^2 C
\right \} ={\rm Det} \Big[ (D^{++})^2\Big]~.
\eea
\end{subequations}
Here $\D_{\rm VM}$ denotes the Faddeev-Popov determinant arising in the vector multiplet model, and the Faddeev-Popov ghosts $\tilde C$ and $C$ are fermionic analytic superfields \cite{Galperin:1985bj,GIOS} of the $\o$-hypermultiplet type \cite{GIKOS}. In accordance with \cite{Faddeev:1967fc}, $\hat{\d } \big[H^{++}  \big]$ does not change if the gauge condition is deformed by 
$D^{++} {\mathfrak V}^{++} \to D^{++} {\mathfrak V}^{++} +\O^{(4)}$, where 
$\O^{(4)}$ is a background analytic superfield. Thus we obtain a more general representation for $\hat{\d } \big[H^{++}  \big]$:
\bea
\hat{\d } \big[H^{++}  \big]&=& \int \big[\mathscr{D} {\mathfrak V}^{++}\big] \,
 \d\big[ D^{++} {\mathfrak V}^{++} -\O^{(4)}\big] \D_{\rm VM}
 \exp \left\{ \ri \int {\rm d}\zeta^{(-4)}\, {\mathfrak V}^{++} H^{++} \right\} ~.
 \label{deltaH}
 \eea

Making use of the identity
\bea
\d\big[ D^{++} {\mathfrak V}^{++} \big]  =  \int \big[\mathscr{D} \r\big] 
\exp \left\{ \ri \int {\rm d}\zeta^{(-4)}\, {\mathfrak V}^{++} D^{++} \r
\right \}~, 
\eea
another representation for $\hat{\d } \big[H^{++}  \big]$ can be obtained, which is:
\bea
\hat{\d } \big[H^{++}  \big]&=& \int 
\big[\mathscr{D} \r\big]  
{\d } \big[H^{++} +D^{++} \r \big]\D_{\rm VM}
~.
\eea

Relation \eqref{gen-delta} defines the generalised delta-function $\hat{\d } \big[H^{++} \big] $ which should be used when trying to apply a modified version of the Faddeev-Poppov quantisation. 
Our next task is to compute
a path integral over the Abelian gauge group \eqref{VM-gauge-tr}
\bea
\int \rd \m_{{\mathfrak U}^{++}}
\hat{\d } \left[H^{++} - \frac{1}{8} (D^+)^2 \L[{\mathfrak U}^{++} ]  - \frac{1}{8} (\bar D^+)^2 \bar \L[{\mathfrak U}^{++} ]  \right]
\eea
First of all, we need to identify a correct integration measure. Given a functional $F[\J , \bar \J]$, let us 
consider an integral 
\bea
I[\J, \bar \J] \propto \int \big[\mathscr{D} {\mathfrak U}^{++}\big] \, F\Big[\J + \ri \L[{\mathfrak U}^{++} ] , 
\bar \J - \ri \bar \L[{\mathfrak U}^{++} ] \Big]
\equiv \int \big[\mathscr{D} {\mathfrak U}^{++}\big] \, 
F\big[\J^{{\mathfrak U}^{++} } , 
\bar \J^{{\mathfrak U}^{++} } 
\big]
~.
\eea
This integral is ill-defined since the integrand is invariant under the gauge transformations \eqref{gauge-sigma}.
Once again, this is the gauge freedom for the Abelian vector multiplet, and it can be fixed by applying the Faddeev-Popov procedure to result with 
\bea
I[\J, \bar \J] &=& \int \rd \m_{{\mathfrak U}^{++}}\, 
F\big[\J^{{\mathfrak U}^{++} } , \bar \J^{{\mathfrak U}^{++} } \big]
~, 
\non \\
&& \rd \m_{{\mathfrak U}^{++}} =
\big[\mathscr{D} {\mathfrak U}^{++}\big] \d\big[ D^{++} {\mathfrak U}^{++} -\U^{(4)}\big] 
\D_{\rm VM}~.
\label{measureU}
\eea
Here $\U^{(4)}$ is a background analytic superfield. 

Now, were are prepared to evaluate 
\bea
(\D_{\rm TM})^{-1} &:=&  \int \rd \m_{{\mathfrak U}^{++}}\, 
\hat{\d } \left[H^{++} \big(  \J^{{\mathfrak U}^{++} } , \bar \J^{{\mathfrak U}^{++} } \big) \right]~,
\eea
where $H^{++} (\J, \bar \J )$ denotes the gauge-fixing condition \eqref{gaugecondition}.
Keeping in mind \eqref{deltaH} and \eqref{measureU}, this can be rewritten as
\bea
(\D_{\rm TM})^{-1} &=& \int \big[\mathscr{D} {\mathfrak V}^{++}\big]  \big[\mathscr{D} {\mathfrak U}^{++}\big]  \d\big[ D^{++} {\mathfrak V}^{++} -\O^{(4)}\big] 
\d\big[ D^{++} {\mathfrak U}^{++} -\U^{(4)}\big] (\D_{\rm VM})^2 \non \\
&&\times 
\exp \left\{ \ri \int {\rm d}\zeta^{(-4)}\, {\mathfrak V}^{++} \Big(H^{++} 
- \frac{1}{8} (D^+)^2 \L[{\mathfrak U}^{++} ]  - \frac{1}{8} (\bar D^+)^2 \bar \L[{\mathfrak U}^{++} ]\Big)\right\} ~.
\label{3.18}
\eea
Making use of the relation \eqref{V--} allows us to rewrite the expression in second line as 
\bea
\exp \left\{ \ri \int {\rm d}\zeta^{(-4)}\, {\mathfrak V}^{++} \Big(H^{++} 
-\frac{1}{16} (D^+)^2 (\bar D^+)^2  {\mathfrak U}^{--} \Big)\right\} ~.
\eea
Next we perform the following shift in the path integral
\begin{subequations}
\bea
{\mathfrak U}^{++} \to {\mathfrak U}^{++} +\frac{\ri}{4} \Box^{-1} \Big( (D^+)^2 \J - 
(\bar D^+)^2 \bar \J \Big)~,
\eea
which implies 
\bea
{\mathfrak U}^{--} \to {\mathfrak U}^{--} +\frac{\ri}{4} \Box^{-1} \Big( (D^-)^2 \J - 
(\bar D^-)^2 \bar \J \Big)~.
\eea
\end{subequations}
We also point out that 
\bea
\frac{1}{16}\int {\rm d}\zeta^{(-4)}\, {\mathfrak V}^{++}
 (D^+)^2 (\bar D^+)^2  {\mathfrak U}^{--} 
 = \int \rd^{4|8}z \rd u_1 \rd u_2 \,\frac{ {\mathfrak V}^{++}(u_1) {\mathfrak U}^{++}(u_2)} 
 {(u^+_1u^+_2)^2} ~,
 \eea
 where $\rd^{4|8}z :=  \rd^4 x \rd^4 \q \rd^4 \bar \q $ is the integration measure of $\cN=2$ Minkowski superspace. 
As a result, \eqref{3.18} turns into 
\bea
(\D_{\rm TM})^{-1} &=& \int \big[\mathscr{D} {\mathfrak V}^{++}\big]  \big[\mathscr{D} {\mathfrak U}^{++}\big]  \d\big[ D^{++} {\mathfrak V}^{++} -\O^{(4)}\big] 
\d\big[ D^{++} {\mathfrak U}^{++} -\U^{(4)}\big] (\D_{\rm VM})^2 \non \\
&&\times 
\exp \left\{ -\ri 
\int \rd^{4|8}z \rd u_1 \rd u_2 \,\frac{ {\mathfrak V}^{++}(1) {\mathfrak U}^{++}(2)} 
 {(u^+_1u^+_2)^2} 
\right\} ~.
\eea
Since the right-hand side is independent of the analytic superfields $\O^{(4)}$ and $\U^{(4)}$, we can integrate over them with a useful weight that we choose
\bea
\label{3.36}
\D_{\rm NK}  \exp \left\{ {\ri}
\int \rd^{4|8}z \rd u_1 \rd u_2 \,\O^{(4)}(1)\, \frac{ (u^-_1 u^-_2)}
 {(u^+_1 u^+_2)^3} \,\U^{(4)}(2) 
\right\} ~,
\eea
with $\D_{\rm NK} $ the Nielsen-Kallosh determinant. The latter is determined by requiring 
\bea
1= \D_{\rm NK}\int \big[\mathscr{D} \O^{(4)}\big]  \big[\mathscr{D} \U^{(4)}\big] \,
  \exp \left\{ {\ri}
\int \rd^{4|8}z \rd u_1 \rd u_2 \,\O^{(4)}(1)\, \frac{ (u^-_1 u^-_2)}
 {(u^+_1 u^+_2)^3} \,\U^{(4)}(2) 
\right\} ~.
\eea
One may evaluate  $\D_{\rm NK}$ following the prescription given in  \cite{Buchbinder:1997ya} (see also \cite{Buchbinder:2001wy} for a review) to end up with 
\bea
\label{NK}
\D_{\rm NK} = \left( \D_{\rm VM} \right)^{-1}  \,  {\rm Det}_{(0,4)} \,
\square ~.
\eea
Here we have introduced the functional determinant ${\rm Det}_{(0,4)}\, \square$ defined by 
\begin{subequations} 
\bea
\left( {\rm Det}_{(0,4)} \,
{\square}{} \right)^{-1} &=&
\int [ \mathscr{D} \r^{(+4)}]\,[ \mathscr{D} \s]
\exp \left\{  \ri \, {\rm tr} \int {\rm d}\zeta^{(-4)} \,
\r^{(+4)} {\square}{} \,\s \right\}~,
\eea
where $\r^{(+4)}$ and $\s$ are unconstrained analytic superfields.
In what follows, we will also need the following result
\bea
\left( {\rm Det}_{(2,2)} \,
{\Box}{} \right)^{-1} &=&
\int [\mathscr{D} U^{++}]\, [ \mathscr{D} V^{++}]\,
\exp \left\{ \ri \, {\rm tr} \int  {\rm d}\zeta^{(-4)} \,
U^{++} \square \,V^{++} \right\}~,  
\eea
\end{subequations}
The result of evaluating $\D_{\rm TM}$ is 
\bea
\D_{\rm TM} = (\D_{\rm NK})^{-1} \,(\D_{\rm VM})^{-2} \, {\rm Det}_{(2,2)} \,{\square}{} 
= (\D_{\rm VM})^{-1} \, 
 \left( {\rm Det}_{(0,4)} \,\square  \right)^{-1} \,
{\rm Det}_{(2,2)} \,\square{} ~.
\eea

At this stage, all prerequisites have been derived in order 
to quantise the tensor multiplet model
\eqref{massless-tensor}. We start with the formal expression 
\bea
\int \big[\mathscr{D} \J  \big] \big[\mathscr{D} \bar \J  \big] \exp\left\{ \ri S[\J, \bar \J] \right\}
\label{3.26}
\eea  
and insert under the integral the following unit
\bea
1&=& \D_{\rm TM} \int \rd \m_{{\mathfrak U}^{++}}\, 
\hat{\d } \left[H^{++} \big(  \J^{{\mathfrak U}^{++} } , \bar \J^{{\mathfrak U}^{++} } \big) \right] \non \\
&=&\D_{\rm TM} \int \rd \m_{{\mathfrak U}^{++}}\, 
\big[\mathscr{D} \r\big]  
{\d } \big[H^{++} +D^{++} \r -\X^{++}\big]\D_{\rm VM}~,
\label{3.27}
\eea
where $\X^{++} $ is a background analytic superfield. By construction, \eqref{3.27} is 
independent of $\X^{++}$. After inserting \eqref{3.27} into \eqref{3.26}, we make the change of variables $\J \to \J^{-{\mathfrak U}^{++} }$. As a result, the infinite group volume $ \int \rd \m_{{\mathfrak U}^{++}}$ factorises and we end up with the partition function
\bea
Z = \int \big[\mathscr{D} \J  \big] \big[\mathscr{D} \bar \J  \big] 
\big[\mathscr{D} \r\big]  
{\d } \big[H^{++} +D^{++} \r -\X^{++}\big] \D_{\rm TM} \D_{\rm VM}
\exp\left\{ \ri S[\J, \bar \J] \right\}~.
\eea
It remains to integrate the right-hand side over $\X^{++}$ with a convenient weight 
\bea
 \exp \left \{ \frac{\ri }{2}   \int {\rm d}\zeta^{(-4)}\, (\X^{++})^2 \right\}~.\non
 \eea
 This leads to 
 \bea
Z = \int \big[\mathscr{D} \J  \big] \big[\mathscr{D} \bar \J  \big] 
\big[\mathscr{D} \r\big]  
\D_{\rm TM} \D_{\rm VM}
\exp\left\{ \frac{\ri}{2} \int \rd^{4|8} z\, \bar \J \J 
+ \frac{\ri}{2} \int {\rm d}\zeta^{(-4)}\,(D^{++} \r)^2  \right\}~.
\eea

Evaluating the path integral over $\J$ and $\bar \J$ gives 
\bea
 \int \big[\mathscr{D} \J  \big] \big[\mathscr{D} \bar \J  \big] 
 \exp\left\{ \frac{\ri}{2} \int \rd^{4|8} z\, \bar \J \J \right\} =
   \left( {\rm Det}\, \cH \right)^{-1/2} 
  ~,
 \eea
 where we have introduced the following operator  
\bea
\mathcal{H} = 
	\begin{pmatrix}
		0 & \bar D^4 \\ 
		D^4 & 0
	\end{pmatrix} ~, \qquad \bar D^4 = \frac{1}{48} \bar D^{ij}\bar D_{ij}~,
	\label{H-operator}
\eea
which acts on the space of chiral/antichiral column vectors $(\J, \bar \J)^{\rm T}$. 
As a result, for the partition function we obtain 
\bea
Z= \Big\{
\left( {\rm Det}\, \cH \right)^{-1/2}   \left({\rm Det}_{(2,2)} \,{\square}{} \right) \, 
\left( {\rm Det}_{(0,4)} \,{\square}{} \right)^{-1} \Big\} 
\Big({\rm Det} \big[ (D^{++})^2\big]\Big)^{-1/2}~.
\eea
The obtained expression simplifies drastically by taking into account the following identity: 
\bea
\left( {\rm Det}\, \cH \right)^{-1/2}   \left({\rm Det}_{(2,2)} \,{\square}{} \right) \, 
\left( {\rm Det}_{(0,4)} \,{\square}{} \right)^{-1} =1~.
\label{3.48}
\eea
This identity may be derived by making use of a change of variables,
$\{ \J, \bar \J , \o , \s \}\to \{ V^{++}, U^{++} \}$, defined by 
\begin{subequations}
\bea
V^{++} &=& D^{++} \o + G^{++} ~,\\
U^{++} &=& D^{++} \s +H^{++}~,
\eea
\end{subequations}
and its inverse. Here $G^{++}$ and $H^{++}$ are defined as in  \eqref{3.7a} and
\eqref{gaugecondition}, respectively. In a supergravity background, 
\eqref{3.48} is expected to be replaced by a topological invariant containing the $\cN=2$ Gauss-Bonnet term \cite{BdeWKL}.

Our final result for the partition function of the tensor multiplet is 
\bea
Z=\Big( {\rm Det}\, \big[ (D^{++})^2\big]\Big)^{-1/2}~.
\eea
This is exactly the partition function of the $\o$ hypermultiplet. Our conclusion will become non-trivial if a supergravity background is turned on.

The main results of this section may be extended to models for a self-interacting tensor multiplet described in \cite{GIO1, GIOS}. This amounts to replacing the free Lagrangian in \eqref{massless-tensor} with an interacting one,  
\bea 
\hf (G^{++})^2 ~\to ~ \cL^{(+4)} \Big( G^{++}, u^\pm \Big)~.
\eea


\section{Concluding comments} \label{section4}

In section \ref{section2} we studied quantum equivalence of the super FT model \eqref{N=1FT} and the nonlinear $\s$-model \eqref{NLSM}. This equivalence may be immediately extended to the more general class of theories defined by replacing the $V^2$ term in \eqref{N=1FT} with a general function of $V$, 
\begin{align}
	S[\chi,\bar{\chi},V] = \hf \text{tr} \int \rd^{4|2}z_+ \, \mathcal{E} \, \chi^\a W_\a 
	+ \hf \text{tr} \int \rd^{4|2}z_- \, \bar{\mathcal{E}} \, \bar{\chi}_\ad \bar{W}^\ad 
	+ \text{tr} \int \rd^{4|4} z \, E \, \mathfrak{F}(V)~,
\end{align}
Specifically, by fixing $\mathfrak{F}(V)$ and repeating the same analysis as in section \ref{section2}, one may show quantum equivalence to the following $\s$-model
\begin{align}
	S[\Phi,\bar{\Phi}] = \text{tr} \int \rd^{4|4} z \, E \,\mathfrak{F}(\re^{\bar{\F}} \re^{\F}) ~.
\end{align}

Additionally, one may instead consider the model obtained by replacing the $V^2$ term in \eqref{N=1FT} with the super Yang-Mills action
\begin{align}
	S_\text{T}[\chi,\bar{\chi},V] = \hf \text{tr} \int \rd^{4|2}z_+ \, \mathcal{E} \, \Big \{\chi^\a W_\a + \frac{1}{g^2}  W^2 \Big \} + \hf \text{tr} \int \rd^{4|2}z_- \, \bar{\mathcal{E}} \, \Big\{ \bar{\chi}_\ad \bar{W}^\ad  +\frac{1}{g^2}  \bar W^2 \Big \}~.
	\label{4.3}
\end{align}
In addition to invariance under \eqref{1stGT}, this model enjoys super Yang-Mills gauge symmetry\footnote{We recall that $L_A B = [A,B]$, for operators $A$ and $B$.}
\begin{align}
	\label{SYMgt}
	\d_\L V = - {\ri} L_V (\L + \bar{\L}) + {\ri} L_V \coth(L_V) (\bar{\L} - \L)~, \qquad \bar{\mathfrak{D}}_\ad \L = 0~.
\end{align}
Further, it is classically topological; it describes trivial dynamics. This property proves to also hold at the quantum level, which may be seen via a slight modification of the analysis of section \ref{section2}. Specifically, it is necessary to incorporate the ghost action corresponding to \eqref{SYMgt} and choose a gauge fermion $\J(\bm{\F})$ such that the gauge $(\mathfrak{D}^2 - 4 \bar{R}) V = (\bar{\mathfrak D}^2 - 4 R) V = 0$ is enforced. With these adjustments, a similar analysis to that conducted in section \ref{section2} leads to $Z=1$ for this theory, implying trivial dynamics. 

An interesting open problem is to extend the results of section \ref{section3} to a curved supergravity background. There are two general superspace approaches to formulate off-shell supergravity-matter systems in four dimensions. The harmonic superspace approach offers powerful prepotential formulations for 
$\cN=2$ supergravity \cite{SUGRA-har1,SUGRA-har2} 
(reviewed in \cite{GIOS, Ivanov:2022vwc}).
The projective superspace approach proves to be ideal for developing covariant geometric formulations for supergravity-matter systems  with eight supercharges
\cite{KLRT-M1,K2008,KT-M2009, KLRT-M2}.
With the advent of $\cN=2$ conformal superspace \cite{ButterN=2}, and its applications to component reduction \cite{Butter:2012xg},
a novel formulation of curved projective superspace has been given in \cite{Butter:2014gha,Butter:2014xua}. 
This approach has also been extended to a novel covariant harmonic superspace framework  in \cite{Butter:2015nza}. For a review of covariant superspace approaches to ${\cal N}=2$ supergravity, see \cite{Kuzenko:2022ajd}. It appears that a combination of harmonic and projective superspace methods is required in order  to extend the results of section \ref{section3} to $\cN=2$ supergravity.

Covariant $\cN=2$ supergravity techniques are expected to be indispensable for computing one-loop effective actions in curved superspace.  For instance, the operator 
\eqref{H-operator} will turn into 
\bea
\mathcal{H} = 
	\begin{pmatrix}
		0 & \bar \D \\ 
		\D & 0
	\end{pmatrix} ~, 
		\label{H-operator-curved}
\eea
where   $\bar{\D}$ denotes the chiral projecting operator \cite{KT-M2009,Muller}
\bea
\bar{\D}
&=&\frac{1}{96} \Big((\cDB^{ij}+16\bar{S}^{ij})\cDB_{ij}
-(\cDB^{\ad\bd}-16\bar{Y}^{\ad\bd})\cDB_{\ad\bd} \Big)
\non\\
&=&\frac{1}{96} \Big(\cDB_{ij}(\cDB^{ij}+16\bar{S}^{ij})
-\cDB_{\ad\bd}(\cDB^{\ad\bd}-16\bar{Y}^{\ad\bd}) \Big)~.
\label{chiral-pr}
\eea
Here $\cD_A= (\cD_a, \cD_\a^i, \bar \cD_i^\ad)$ are the covariant derivatives in curved superspace, 
$\cDB^{\ad\bd}:=\cDB^{(\ad}_k\cDB^{\bd)k}$, and $\bar{S}^{ij}$ and $\bar{Y}^{\ad\bd}$ are certain torsion tensors, see \cite{Kuzenko:2022ajd} for a review.
The main properties of $\bar \D$ include the following:
for any scalar $U$, it holds that 
\begin{subequations} 
\bea
{\bar \cD}^{\ad}_i \bar{\D} U &=&0~, \\
\int \rd^4 x \,{\rm d}^4\q\,{\rm d}^4{\bar \q}\,E\, U
&=& \int {\rm d}^4x \,{\rm d}^4 \q \, \cE \, \bar{\D} U ~,
\label{chiralproj1} 
\eea
\end{subequations}
where $E$ and $\cE$ denote, respectively, the full superspace and the chiral subspace densities.
Operator $\bar \D \D $ is a covariantly chiral fourth-order operator when acting on the space of covariantly chiral scalar superfields,
\bea
 \bar \cD^\ad_i \J=0 \quad \implies \quad \bar \D \D \J = \Big\{ (\cD^a \cD_a)^2 + \dots \Big\} \J~.
\eea

An $\cN=2$ locally supersymmetric analogue of the topological model \eqref{4.3} is
\bea
S=
\hf   \int {\rm d}^4 x \rd^4\q  \, \cE\, W\big( \J  +\frac{1}{g^2} W\big)
+{\rm c.c.}  ~,
\eea
where the prepotential $\J$ for the tensor multiplet is now covariantly chiral, 
and chiral field strength $W$ for the vector multiplet obeys the constraints
\bea
\cDB^\ad_i W= 0~, \qquad
 \Big(\cD^{\a(i}\cD_\a^{j)}+4S^{ij}\Big) W= 
\Big(\cDB_\ad{}^{(i}\cDB^{j) \ad}+4\bar{S}^{ij}\Big)\bar{W} ~.
\eea
This model respects the gauge symmetries of both tensor and vector multiplets. It would be interesting to compute the partition function for the model since its logarithmically divergent part should contain  
the $\cN=2$ Gauss-Bonnet term \cite{BdeWKL}.

In the supersymmetric literature, the BV procedure has been employed 
\cite{Grisaru:1997hf, Penati:1997pm, Grassi:2000it, TartaglinoMazzucchelli:2004vt}
to quantise theories involving the massless non-minimal scalar multiplet \cite{GS81} described by a complex linear superfield. These are reducible gauge theories of infinite stage of reducibility, which is similar to the Green-Schwarz superstring. Strictly speaking, the BV approach is directly applicable to quantise finitely reducible gauge theories. However, a way out was found in Refs. \cite{Grisaru:1997hf, Penati:1997pm} and their extensions \cite{Grassi:2000it, TartaglinoMazzucchelli:2004vt}.
Another family of reducible supersymmetric theories are the massless higher-spin supermultiplets in AdS$_4$ \cite{KS94, Buchbinder:2018nkp}. When realised in terms of unconstrained superfield prepotentials introduced in \cite{KS94, KPS,KS, Buchbinder:2018nkp}, their stage of reducibility is finite or infinite depending on the superspin and the type of the formulation. Lagrangian quantisation of the higher-spin supermultiplets in AdS$_4$ \cite{KS94} was carried out in \cite{Buchbinder:1995ez} using a special simplification of the BV procedure in reduction coordinates (transversal irreducible superfields) for a general quadratic action. It would be interesting to come back to the problem of BV quantisation of the supersymmetric higher-spin gauge models in AdS$_4$ in terms of unconstrained superfields. 
\\

\noindent
{\bf Acknowledgements:}  
This work was supported in part by the Australian Research Council, project No. DP230101629. Since May 1, 2024, the work of ER is supported by the Brian Dunlop Physics Fellowship. We acknowledge the kind hospitality and financial support extended to us at the MATRIX Program ``New Deformations of Quantum Field and Gravity Theories'' (between 22 January and 2 February, 2024) where part of this work was performed.

\appendix 

\section{Rudiments of harmonic superspace}
\label{appendixA}

The $\cN=2$ harmonic superspace
${\Bbb R}^{4|8} \times S^2$
\cite{GIKOS,GIOS} extends conventional $\cN=2$ superspace ${\mathbb R}^{4|8} $ 
(paramerised by coordinates $z^A= (x^a , \q^\a_i , {\bar \q}_\ad^i)$, with $i =\hat{1},  \hat{2}$) 
by the two-sphere $S^2 = \sSU(2)/\sU(1)$
parametrised by harmonics
elements
\bea
({u_i}^-\,,\,{u_i}^+) \in \sSU(2)~, \quad
u^+_i = \ve_{ij}u^{+j}~, \quad \overline{u^{+i}} = u^-_i~,
\quad u^{+i}u_i^- = 1 ~.
\eea
Instead of using the standard basis for spinor covariant derivatives $D^i_\a$ and ${\bar D}^\ad_i$, with $i =\hat{1},  \hat{2}$, which obey the anti-commutation relations
\bea
\{ D^i_\a , D^j_\b \} 
= \{ {\bar D}^\ad_i , {\bar D}^\bd_j \} =0~, 
\qquad 
\{ D^i_\a , {\bar D}_{\bd j} \} = -2{\rm i}\,
\d^i_j \,(\s^c)_{\a \bd} \, \pa_c~,
\eea 
a new harmonic-dependent basis can be introduced by the rule
\bea
D^\pm_\a = D^i_\a \,u^\pm_i~, \qquad 
{\bar D}^\pm_\ad = {\bar D}^i_\ad \,u^\pm_i~.
\label{spinor-der}
\eea
The operators $D^+_\a $ and $\bar D^+_\ad $ strictly anticommute, and therefore one can introduce analytic superfields of $\sU(1)$ charge $n$, $\vf^{(n)}(z, u)$,  with the properties: 
\begin{subequations}
\bea
&D^+_\a \vf^{(n)} =0~, \qquad \bar D^+_\ad \vf^{(n)} =0;\\
&\vf^{(n)}(z, {\rm e}^{ {\rm i}\a}\, u^+, {\rm e}^{ -{\rm i}\a}\,u^-) 
= {\rm e}^{ {\rm i}n\a} \,\vf^{(n)}(z,u^+, u^-)~.
\eea
\end{subequations}
Such superfields live on the so-called analytic subspace of the harmonic superspace.
When working in the harmonic-superspace approach, of special significance are the operators 
\bea
D^{++}=u^{+ i}\frac{\partial}{\partial u^{- i}} ~, \qquad
D^{--}=u^{- i}\frac{\partial}{\partial u^{+ i}} ~.
\eea

Using the fourth-order operators 
\bea 
(D^+)^4 = {1 \over 16} (D^+)^2 ({\bar D}^+)^2 ~, 
\qquad  
(D^-)^4 = {1 \over 16} (D^-)^2 ({\bar D}^-)^2 ~, 
\label{A.5}
\eea
integration over the analytic subspace is defined by
\bea
\int {\rm d}\z^{(-4)} \, L^{(+4)} = 
\int {\rm d}^4 x \int {\rm d}u\, (D^-)^4 L^{(+4)}~, 
\qquad 
D^+_\a L^{(+4)} =
{\bar D}^+_\ad L^{(+4)} =0~.
\eea
Additionally, integration over the group manifold 
$\sSU(2)$ is defined according to \cite{GIKOS}
\bea 
 \int {\rm d}u \, 1 = 1~\qquad 
 \int {\rm d}u \, u^+_{(i_1} \cdots u^+_{i_n}\,
u^-_{j_1} \cdots u^-_{j_m)} =0~, 
\quad n+m >0~.
\eea

\section{BV quantisation of $\cN=2$ gauge theories}
\label{appendixB}

This appendix is devoted to the BV quantisation of two $\cN=2$ supersymmetric gauge theories in harmonic superspace. Specifically, we will consider the Abelian vector\footnote{Quantisation of $\cN=2$ super Yang-Mills theory was performed in \cite{Buchbinder:1997ya} in the framework of the background field method. For an Abelian gauge group, its results are in agreement with the ones presented below.} and tensor multiplets. While the latter was quantised in section \ref{section3} via a modified Faddeev-Popov procedure, we hope that this technical appendix is of use to the reader unfamiliar with the formalism.

\subsection{The vector multiplet}
The $\cN=2$ vector multiplet is known to be described by a real analytic superfield $V^{++}$
\begin{align}
	D^{+}_\a V^{++} = 0 ~, \qquad \bar{D}^{+}_\ad V^{++} = 0~,
\end{align}
which is defined modulo the following Abelian gauge transformations
\begin{align}
	\label{VMGT}
	\d_\l V^{++} = - D^{++} \l ~, \qquad D^+_\a \l = 0 ~, \quad \bar{D}^+_\ad \l = 0 ~.
\end{align}
It may then be shown that the following action for $V^{++}$
\begin{align}
	\label{B.2}
	S_\text{VM}[V^{++}] = \hf \int \rd^{4|8} z {\rm d}u_1 {\rm d} u_2 \, \frac{V^{++}(1) V^{++}(2)}{(u_1^+ u_2^+)^2}~,
\end{align}
is inert under \eqref{VMGT}. Below, we will quantise this model by employing the BV formalism.

In accordance with eq. \eqref{2.21}, the minimal solution to the master equation \eqref{mastereq} is
\begin{align}
	S[\bm{\F}_\text{min},\bm{\F}^*_\text{min}] = S_\text{VM}[V^{++}] + \int {\rm d}\zeta^{(-4)}\, V^{* ++} D^{++} C~, \qquad \bm{\F}_\text{min} = \{ V^{++}, C\}~,
\end{align}
where $C$ is a fermionic analytic superfield; $D_\a^+ C = 0$, $\bar{D}_\ad^+ C = 0$. To eliminate antifields we append to $\bm{\F}_\text{min}$ the following pair of non-minimal superfields:
\begin{align}
	(\tilde{C},\pi)~, \quad D_\a^+ C = D_\a^+ \pi = 0 ~, \quad \bar{D}^+_\ad C = \bar{D}_\ad^+ \pi = 0~, \quad \e(\tilde{C}) = \e(\pi) + 1 = 1~.
\end{align}
Then, the following functional\footnote{Given an analytic superfield of charge $n$, its corresponding antifield carries charge $4-n$.}
\begin{align}
	\label{B.6}
	S[\bm{\F},\bm{\F}^*] = S[\bm{\F}_\text{min},\bm{\F}^*_\text{min}] + \int {\rm d}\zeta^{(-4)}\, \tilde{C}^{*(4)} \pi~,
\end{align}
is also a solution of the master equation and satisfies $\D S[\bm{\F},\bm{\F}^*] = 0$. Now, we choose the gauge fermion to be
\begin{align}
	\label{B.7}
	\J[\bm{\F};\L^{(4)}] = \int {\rm d}\zeta^{(-4)}\, \Big ( \tilde{C} D^{++} V^{++} - \tilde{C} \L^{(4)} \Big )~, \qquad D_\a^{+} \L^{(4)} = 0 ~, \quad \bar{D}_\ad^{+} \L^{(4)} = 0~,
\end{align}
where $\L^{(4)}$ is a background analytic superfield. Eliminating antifields in \eqref{B.6} in accordance with \eqref{B.7} yields the following
\begin{align}
	S\Big[\bm{\F},\frac{\d \J}{\d \bm{\F}}\Big] = S_\text{VM}[V^{++}] + \int {\rm d}\zeta^{(-4)}\, (D^{++} V^{++} - \L^{(4)}) \pi ~.
\end{align}

This analysis indicates that the in-out vacuum amplitude for \eqref{B.2} is
\begin{align}
	Z = \int \big[\mathscr{D} V^{++}  \big] \,\d(D^{++} V^{++} - \L^{(4)}) \exp \Big(\ri S_\text{VM}[V^{++}] \Big) ~.
\end{align}
Since the right-hand side is independent of the analytic superfield $\L^{(4)}$, we can integrate over it with the weight
\bea
\tilde{\D}_{\rm NK}  \exp \left\{ \frac{\ri}{2 \a}
\int \rd^{4|8}z \rd u_1 \rd u_2 \,\L^{(4)}(1)\, \frac{ (u^-_1 u^-_2)}
{(u^+_1 u^+_2)^3} \,\L^{(4)}(2) 
\right\} ~, \qquad \a \in \mathbb{R} - \{0\}~,
\eea
where $\tilde{\D}_{\rm NK} $ is a Nielsen-Kallosh determinant. The latter is determined by requiring 
\bea
1= \tilde{\D}_{\rm NK}\int \big[\mathscr{D} \L^{(4)}\big] \,
\exp \left\{ {\ri}
\int \rd^{4|8}z \rd u_1 \rd u_2 \,\L^{(4)}(1)\, \frac{ (u^-_1 u^-_2)}
{(u^+_1 u^+_2)^3} \,\L^{(4)}(2) 
\right\} ~.
\eea
In accordance with the analysis of \cite{Buchbinder:1997ya}, it is given by
\begin{align}
	\tilde{\D}_{\rm NK} = \Big(\text{Det} [(D^{++})^2] \Big)^{- \hf}\, (\text{Det}_{(0,4)} \Box )^{\hf}~.
\end{align}

A routine manipulation of the resulting path integral leads to 
\begin{align}
	 Z&= \int \big[\mathscr{D} V^{++}  \big] (\text{Det} [(D^{++})^2])^{\hf}\, (\text{Det}_{(0,4)} \Box )^{\hf}  \non \\
	 &\qquad \qquad \times \exp\Big(\ri \int \rd^{4|8}z \rd u_1 \rd u_2\,  \hf \Big(1 + \frac{1}{\a}\Big) \frac{V^{++}(1)V^{++}(2)}{(u_1^+ u_2^+)^2} - \frac{\ri}{2 \a} \int {\rm d}\zeta^{(-4)}\, V^{++} \Box V^{++} \Big] \Big) ~.
\end{align}
Finally, fixing $\a = -1$ yields Feynman gauge and we obtain
\begin{align}
	Z = \Big(\text{Det} [(D^{++})^2]\Big)^{\hf}\, (\text{Det}_{(0,4)} \Box )^{\hf} \, (\text{Det}_{(2,2)} \Box )^{-\hf}~.
\end{align}

\subsection{The tensor multiplet}
We recall that the $\cN=2$ tensor multiplet is described by action \eqref{massless-tensor}, which we reiterate for convenience
\bea
S_\text{TM}[\J, \bar \J]=  \hf 
\int {\rm d}\zeta^{(-4)}\,(G^{++})^2~, \qquad G^{++}(z,u)= \frac{1}{8} (D^+)^2 \J (z)
+\frac{1}{8} ({\bar D}^+)^2 {\bar \J} (z) ~,
\label{massless-tensor2}
\eea
where $\J$ is a chiral superfield; $\bar{D}_\ad^i \J = 0$. This action is invariant under the Abelian gauge transformations \eqref{3.3}, which can be expressed in terms of a real analytic superfield \eqref{reduced}
\begin{align}
	\label{gt1}
	\d_{\mathfrak U} \Psi = \frac{\ri}{4} \int {\rm d}u \, 
	({\bar D}^-)^2 {\mathfrak U}^{++}~, \qquad D^+_\a {\mathfrak U}^{++} =0~,
	\quad \bar D^+_\ad {\mathfrak U}^{++} =0~.
\end{align}
This gauge symmetry is reducible as \eqref{gt1} is inert under
\begin{align}
	\label{gt2}
	\d_\s{\mathfrak U}^{++} = D^{++} \s ~, \qquad D^+_\a \s =0~,
	\quad \bar D^+_\ad \s =0~.
\end{align}
Below, we will employ the BV scheme to quantise this model and provide an alternative proof of quantum equivalence to the $\o$ hypermultiplet model \eqref{3.16} to the one given in section \ref{section3.3}.

Making use of \eqref{gt1} and \eqref{gt2} in conjunction with \eqref{2.21}, we obtain the minimal solution to the master equation \eqref{mastereq}
\begin{align}
	S[\bm{\F}_\text{min},\bm{\F}^*_\text{min}] &= S_\text{TM}[\J,\bar{\J}] + \frac{\ri}{4} \bigg \{ \int \rd^{4|4}z_+ \rd u \, \J^* (\bar{D}^-)^2 C^{++} - \int \rd^{4|4}z_- \rd u \, \bar{\J}^* (D^-)^2 C^{++} \bigg \} \non \\
	& \quad+ \int {\rm d}\zeta^{(-4)}\, C^{* ++} D^{++} \eta~, 
\end{align}
where $\bm{\F}_\text{min} = \{ \J,\bar{\J},C^{++},\eta\}$.\footnote{It should be emphasised that the ghosts $C^{++}$ and $\eta$ are both analytic superfields.} To eliminate antifields we extend $\bm{\F}_\text{min}$ by the following pairs of non-minimal analytic superfields:
\begin{subequations}
	\label{nonminimal2}
	\begin{align}
		(i):& \qquad (\tilde{C}^{++},\pi^{++})~, \quad \e(\tilde{C}^{++}) = \e(\pi^{++}) + 1 = 1~, \\
		(ii):& \qquad (\tilde{\eta},\tilde{\pi})~, \quad \e(\tilde{\eta}) = \e(\tilde{\pi}) - 1 = 0~, \\
		(iii):& \qquad ({\eta}',{\pi}')~, \quad \e({\eta}') = \e({\pi}') - 1 = 0~. 
	\end{align}
\end{subequations}
Then, the following functional
\begin{align}
	\label{B.20}
	S[\bm{\F},\bm{\F}^*] = S[\bm{\F}_\text{min},\bm{\F}^*_\text{min}] + \int {\rm d}\zeta^{(-4)}\, \Big ( \tilde{C}^{*++} \pi^{++} + \tilde{\eta}^{*(4)} \tilde{\pi} + \eta'^{*(4)} \pi' \Big )~,
\end{align}
is also a solution of the master equation and satisfies $\D S[\bm{\F},\bm{\F}^*] = 0$. We choose the gauge fermion to be\footnote{Here we denote the gauge fermion by $\J_\text{GF}$ to distinguish it from the tensor multiplet's prepotential.}
\begin{align}
	\label{B.21}
	\J_\text{GF}[\bm{\F};\O^{(4)},\U^{(4)}] &= \int {\rm d}\zeta^{(-4)}\, \Big ( \tilde{C}^{++} H^{++} - \tilde{\eta} D^{++} C^{++} + D^{++} \tilde{C} \eta' \non \\
	&\qquad \qquad \qquad\qquad \qquad- \frac{1}{2m} \tilde{C}^{++} \pi^{++} + \O^{(4)} \tilde{\eta} - \U^{(4)} \eta' \Big ) ~,
\end{align}
where $H^{++}$ is given by \eqref{gaugecondition} while $\O^{(4)}$ and $\U^{(4)}$ are fermionic background analytic superfields. Eliminating antifields in \eqref{B.20} in accordance with \eqref{B.21} yields the following
\begin{align}
	S\Big[\bm{\F},\frac{\d \J_{\text{GF}}}{\d \bm{\F}}\Big] &= S_\text{TM}[\J,\bar{\J}] - \int \rd^{4|8}z \rd u_1 \rd u_2\,  \frac{\tilde{C}^{++}(1)C^{++}(2)}{(u_1^+ u_2^+)^2} + \int {\rm d}\zeta^{(-4)}\, \Big ( D^{++} \tilde{\eta} D^{++} \eta \non \\
	&  + \Big(H^{++} - D^{++} \eta' - \frac{1}{2m} \pi^{++}\Big) \pi^{++} + \Big(\O^{(4)} - D^{++} C^{++}\Big) \tilde{\pi} + \Big(D^{++} \tilde{C}^{++} - \U^{(4)} \Big) \pi' \Big)~.
\end{align}

This analysis indicates that the in-out vacuum amplitude for \eqref{B.2} is
\begin{align}
	Z &= \int \big[\mathscr{D} \bar{\J}  \big] \big[\mathscr{D} {\J}  \big] \big[\mathscr{D} \tilde{C}  \big] \big[\mathscr{D} {C}  \big] \big[\mathscr{D} \tilde{\eta}  \big] \big[\mathscr{D} {\eta}  \big] \big[\mathscr{D} \eta' \big] \,\d(D^{++} \tilde{C}^{++} - \O^{(4)}) \d(D^{++} {C}^{++} - \U^{(4)}) \non \\
	&\qquad \times \exp \Big(\ri \Big \{ S_\text{TM}[\J,\bar{\J}] -  \int \rd^{4|8}z \rd u_1 \rd u_2\,  \frac{\tilde{C}^{++}(1)C^{++}(2)}{(u_1^+ u_2^+)^2} + \int {\rm d}\zeta^{(-4)}\, \Big[-\tilde{\eta} (D^{++})^2 \eta \non \\
	&\qquad \qquad \qquad  - \frac{m}{2} \eta' (D^{++})^2 \eta' + \frac{m}{2} (H^{++})^2 \Big] \Big \} \Big)~,
\end{align}
where we have integrated out $\pi^{++}$, $\tilde{\pi}$ and $\pi'$. 
Since the right-hand side is independent of the background analytic superfields $\O^{(4)}$ and $\U^{(4)}$, we can integrate over them with the weight
\bea
\label{B.24}
({\D}_{\rm NK})^{-1}  \exp \left\{ \frac{\ri}{n}
\int \rd^{4|8}z \rd u_1 \rd u_2 \,\O^{(4)}(1)\, \frac{ (u^-_1 u^-_2)}
{(u^+_1 u^+_2)^3} \,\U^{(4)}(2) 
\right\} ~, \qquad n \in \mathbb{R} - \{0\}~,
\eea
where ${\D}_{\rm NK} $ is the Nielsen-Kallosh determinant \eqref{NK}.\footnote{It should be noted that the background fields in \eqref{B.24} are of opposite statistics to those of \eqref{3.36}. This is why their Nielsen-Kallosh determinants are inverse to each other.}  The resulting path integral is
\begin{align}
	\label{B.25}	
	Z &= \int \big[\mathscr{D} \bar{\J}  \big] \big[\mathscr{D} {\J}  \big] \big[\mathscr{D} \tilde{C}  \big] \big[\mathscr{D} {C}  \big] \big[\mathscr{D} \tilde{\eta}  \big] \big[\mathscr{D} {\eta}  \big] \big[\mathscr{D} \eta' \big] \, ({\D}_{\rm NK})^{-1} \exp \Big(\ri \Big \{ S_\text{TM}[\J,\bar{\J}]  \non \\
	&\qquad +  \int \rd^{4|8}z \rd u_1 \rd u_2\,  \Big(\frac{1}{n} - 1\Big) \frac{\tilde{C}^{++}(1)C^{++}(2)}{(u_1^+ u_2^+)^2}  + \frac{1}{2 n} \int {\rm d}\zeta^{(-4)}\, \tilde{C}^{++} \Box C^{++} \non \\
	&\qquad \qquad + \int {\rm d}\zeta^{(-4)}\, \Big[-\tilde{\eta} (D^{++})^2 \eta - \frac{m}{2} \eta' (D^{++})^2 \eta' + \frac{m}{2} (H^{++})^2 \Big] \Big \} \Big )~.
\end{align}

Choosing the gauge $m=n=1$ leads to significant simplifications in \eqref{B.25} and allows one to readily perform the path integral over all remaining fields. This leads to
\begin{align}
	Z&= ({\D}_{\rm NK})^{-1}\left( {\rm Det}\, \cH \right)^{-1/2} ({\rm Det}_{(2,2)} \,{\square}{} ) \, 
	\Big({\rm Det} \big[ (D^{++})^2\big]\Big)^{-3/2} \non \\
	&= ( {\rm Det}\, \cH )^{-1/2} ({\rm Det}_{(2,2)} \,{\square}{} ) \, 
		( {\rm Det}_{(0,4)} \,{\square}{} )^{-1} \,
	\Big({\rm Det} \big[ (D^{++})^2\big]\Big)^{-1/2}~, \non \\
	&= \Big(\text{Det} [(D^{++})^2]\Big)^{-\hf}~,
\end{align}
where we have made use of eq. \eqref{3.48} in the final line. This result corroborates the quantum equivalence of the tensor and $\o$ hypermultiplet demonstrated in section \ref{section3.3}.

\begin{footnotesize}

\end{footnotesize}

\end{document}